\documentclass{PoS}

\newcommand{\ee}{e^{+}e^{-}}

\newcommand{\ec}{\eta_{\,c}}
\newcommand{\hc}{h_{c}}

\newcommand{\etap}{\eta^{\prime}}
\newcommand{\etac}{\eta_{\, c}}
\newcommand{\psip}{\psi^{\prime}}

\newcommand{\pipi}{\pi^{+}\pi^{-}}
\newcommand{\piz}{\pi^{0}}

\newcommand{\DDbar}{D\bar{D}}
\newcommand{\ccbar}{c\bar{c}}
\newcommand{\bbbar}{b\bar{b}}
\newcommand{\qqbar}{q\bar{q}}
\newcommand{\uubar}{u\bar{u}}
\newcommand{\ddbar}{d\bar{d}}
\newcommand{\ssbar}{s\bar{s}}
\newcommand{\ppbar}{p\bar{p}}

\newcommand{\rt}{\rightarrow}

\newcommand{\etal}{\em et al.}
\newcommand{\jpsi}{J/\psi}
\newcommand{\azero}{a_{\, 0}(980)}

\newcommand{\fzero}{f_{\,0}(980)}

\title{News from BESIII}

\ShortTitle{News from BESIII}

\author{Stephen Lars Olsen\\
        Department of Physics \& Astronomy, Seoul National University\\
        Gwanak-gu, Seoul, 151-747, KOREA\\
        E-mail: \email{solsen@hep1.snu.ac.kr}}

\author{Representing the BESIII Collaboration\\
}
\abstract{BESIII is a new state-of-the-art 4$\pi$ detector at the recently upgraded BEPCII two-ring $\ee$ 
          collider at the Institute of High Energy Physics in Beijing.  It has been in operation for
          three years, during which time it has collected the world's largest data samples of $\jpsi$,
          $\psip$ and $\psi(3770)$ decays.  These data are being used to make a variety of interesting
          and unique studies of light-hadron spectroscopy, precision charmonium physics and high-statistics
          measurements of $D$ meson decays.   Results that I describe in this report include studies of
          $\azero$-$\fzero$ mixing, an observation of a large isospin-violation in $\eta(1405)\rt\piz\fzero$
          decays, some puzzles in $\jpsi$ and $\psip$ decays to light hadrons, the observation of
          two glueball candidate states in radiative $\jpsi\rt\gamma\pipi\etap$ decays and some
          recent precision measurements of $\ec$ and $\hc$ lineshapes.}

\FullConference{50th International Winter Meeting on Nuclear Physics - Bormio2012,\\
		23-27 January 2012\\
		Bormio, Italy}

\begin{document}

\section{Introduction}

The BES experimental program dates back to late 1989 when operation of the
Beijing Electron Positron Collider (BEPC) and the Beijing Electron
Spectrometer (BES) first started.  BEPC was a single-ring $\ee$ collider that operated in the
$\tau$-charm threshold energy region between (about) 2.5~and~4.5~GeV with
a luminosity of $\sim 10^{31}$cm$^{-2}$s$^{-1}$.  Among the early
successes included a precise measurement of the mass of the $\tau$ lepton~\cite{tau-mass}
that not only improved on the precision of previous measurements by an order-of-magnitude,
but also showed that the existing world avearge value was high by about two standard
deviations.   Another key result was the precise measurement of the total cross section
for $\ee$ annihilation into hadrons over the accessible center of mass (c.m.) energy
range~\cite{bes_R}.
The precision of these measurements lead to a substantially improved evaulation of
the electromagnetic coupling constant extrapolated to the $Z$-boson
mass peak, $\alpha_{QED}(M^2_Z)$, which resulted in a significant
$\sim$30\% increase in the Standard
Model (SM) predicted value for the Higgs' boson mass~\cite{Higgs-mass}.

In the late 1990s, the BES detector was upgraded to the BESII detector
and this produced a number of interesting results including even
more precise cross section measurements~\cite{bes_R2} and the discovery of a number
of new hadron states, including the $\sigma$~\cite{bes_sigma} and
$\kappa$~\cite{bes_kappa} scalar resonances and a still-unexplained
subthreshold $\ppbar$ resonance produced in radiative $\jpsi\rt\gamma\ppbar$
decays~\cite{bes_x1860}.

Between 2005 and 2008, BEPC was replaced by BEPCII, a two-ring $\ee$ collider
with a hundred-fold increase in luminosity, and  the BESII detector was completely removed and
replaced by BESIII, a state-of-the-art detector built around a 1~T superconducting
solenoid that contains a cylindrical drift chamber, a double-layer barrel of
scintillation counters for time-of-flight measurements, and a nearly $4\pi$ array of 
6240 CsI(Tl) crystals for electromagnetic calorimetry.  The magnet's iron flux-return  
yoke is instrumented with a nine-layer RPC muon identification system.
BEPCII operations started in summer 2008 and since then
the luminosity has been continuously improving; now it is
$\sim 6\times 10^{32}$cm$^{-2}$s$^{-1}$, quite near the  $10^{33}$ design value.
The BESIII detector performance is excellent: the charged particle momentum
resolution is $\delta p/p\simeq 0.5$\%; the $\gamma$ energy resolution is 
2.5\% at $E_{\gamma}=1$~GeV; the 6\% resolution  $dE/dx$  measurements in the drift chamber 
plus the  $\sim$80~ps resolution time-of-flight
measurements is sufficient
to identify charged particles over the entire momentum range of interest.

The BESIII experimental program addresses issues in light hadron physics,
charmonium spectroscopy and decays, $D$ and $D_s$ meson decays, and 
numerous topics in QCD and $\tau$-lepton physics.  To date, BESIII
has accumulated data samples corresponding to 225M $\jpsi$ decays,
106M $\psip$ decays and 2.9~fb$^{-1}$ at the peak of the $\psi(3770)$
resonance, which decays to $\DDbar$ meson pairs nearly 100\% of the time.
These are all world's-largest data samples at these c.m. energies
and the $\jpsi$ sample is the first ever to be collected in
a high quality detector like BESIII.
In this talk I review some recent results on hadron physics that
have been generated from these data samples.

\section{Some issues in light hadron physics}
In the original quark model, first proposed by
Gell-Mann~\cite{gellmann} and Zweig~\cite{zweig}
in 1964,  mesons are comprised of quark-antiquark ($\qqbar$) pairs and
baryons are three-quark ($qqq$) triplets, a picture that accurately classified
the properties of all of the hadronic particles and resonances known at the time. 
Its phenomenal success at predicting virtually all of the detailed
properties of the $\ccbar$ charmonium and $\bbbar$ bottomonium
states that were subsequently discovered in the 1970's led to 
the near unanimous agreement that, in spite of the fact that
an individual quark had never been seen, quarks are real objects
and the quark-antiquark mesons and quark-quark-quark baryons
are the basic combinations that form hadrons.  

In the 1970's, this simple quark model was superseded by
Quantum Chromodynamics (QCD), which identified the underlying reason
for these rules was that $\qqbar$ pair and $qqq$ triplet combinations
are color-singlet representations of the color $SU(3)$ group that is
fundamental to the theory.   Somewhat suprisingly, the 
mesons-are-$\qqbar$ and baryons-are-$qqq$
prescriptions still adequately describe the hadronic particle spectrum
despite the existence of a number of other color-singlet quark and gluon
combinations that are possible in QCD.  These additional combinations include
five-quark ($qqq\qqbar$) ``pentaquark'' baryons, four-quark ($\qqbar\qqbar)$
tetraquark mesons, and mesons formed from valence gluons: either $\qqbar$-$g$
hybrid mesons or $g$-$g$ glueballs.   However, considerable experimental efforts
at searching for color singlet $qqq\bar{q}q$ pentaquark baryons have still
not established any unambiguous candidate for five-quark states~\cite{trilling}, 
and, although a few candidates for non-$\qqbar$ light-hadron meson resonances
have been reported~\cite{e852}, none have been generally accepted as
established by the hadron physics community~\cite{barnes_1}.

\begin{figure}
\mbox{
  \includegraphics[height=0.4\textwidth,width=0.5\textwidth]{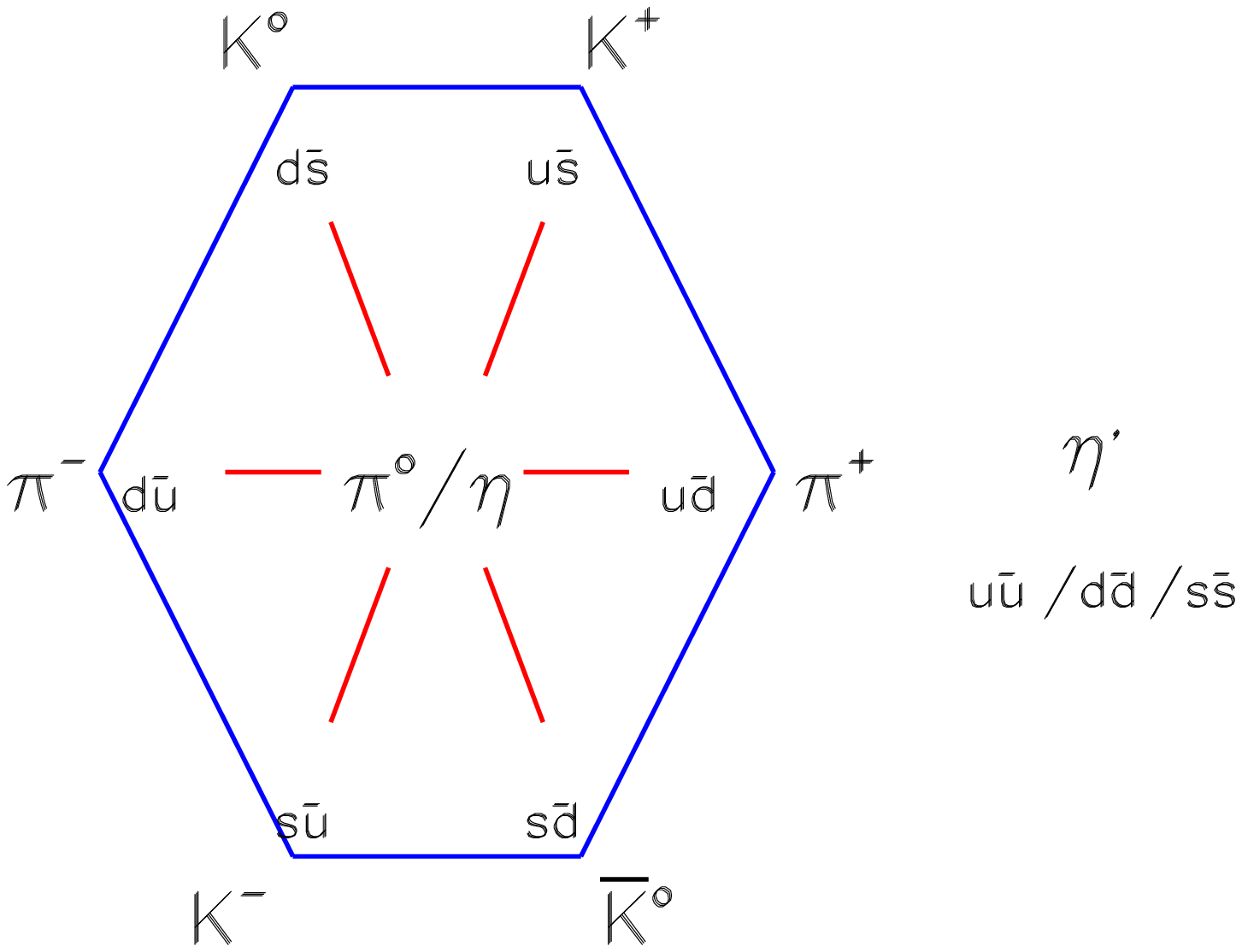}
}
\mbox{
  \includegraphics[height=0.4\textwidth,width=0.5\textwidth]{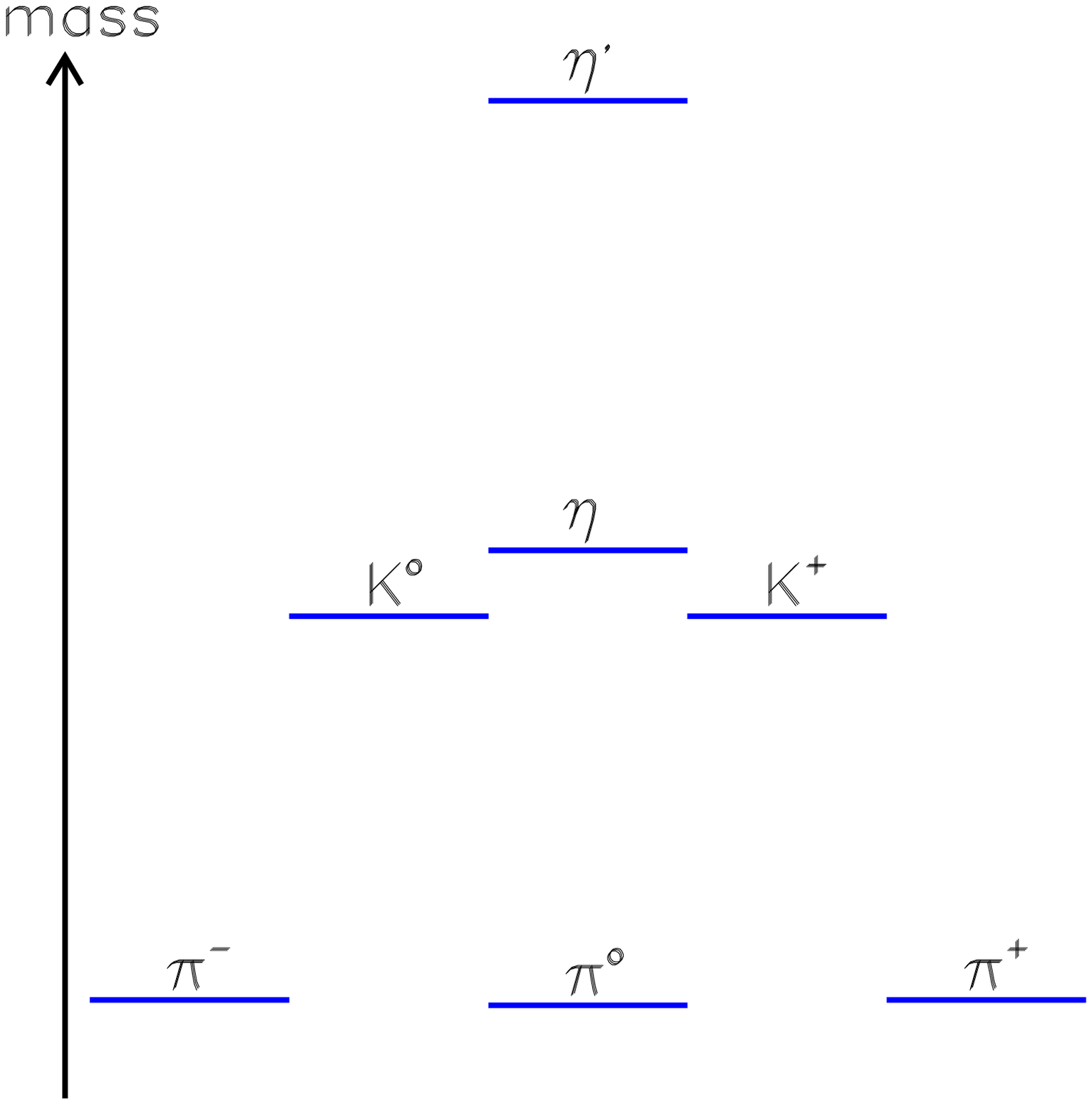}
}
\caption{{\bf Left:} The pseudoscalar meson nonet.
The $\piz$ meson is a $(\uubar - \ddbar)/\sqrt{2}$ and the $\eta$ and
$\etap$ mesons are $\uubar$ $\ddbar$ $\ssbar$ mixtures described in the text.
{\bf Right:} the pseudoscalar
meson mass hierarchy.} 
\label{fig:pseudoscalar_nonet}
\end{figure}

A consequence of the $\qqbar$ mesons and $qqq$ baryon rules
are that light mesons\footnote{{\it i.e.}, mesons formed from only $u$, $d$ \& $s$
quarks.} all come in flavor-$SU(3)$ nonets and baryons come in either
nonets and decuplets.
In any given multiplet, all of the particles would  have the same mass
in the limit where the $u$, $d$ and $s$ quark masses are equal. However,
since the $s$-quark is in fact substantially more massive than its $u$ and $d$
counterparts, this symmetry is broken, and particles containing $s$ quarks are
heavier than their non-strange multiplet partners.  This is illustrated in
Fig.~\ref{fig:pseudoscalar_nonet}, where the left panel indicates the
quark content of the mesons and the right panel shows the meson mass hierarchy. 
The isospin triplet pions have no $\ssbar$ content and are the lightest.
The $\eta$ and $\etap$ mesons are $\uubar$ $\ddbar$ $\ssbar$ mixtures:
\begin{eqnarray}
\eta & = & \cos\phi_P\frac{|\uubar >+|\ddbar >}{\sqrt{2}} - \sin\phi_P |\ssbar >\\
\etap & = &\sin\phi_P\frac{|\uubar >+|\ddbar >}{\sqrt{2}} + \cos\phi_P |\ssbar >,
\label{eq:eta-mix}
\end{eqnarray}
where $\phi_P\simeq 38^{\circ}$~\cite{scadron}.  With this mixing angle, the
$\ssbar$ component accounts for about 35\% of the $\eta$'s wave function
and about 46\% of the $\etap$'s wave function~\cite{okubo}; this large
$\ssbar$ content of the $\eta$ and $\etap$ is reflected in their relatively
high masses.   Similar hierarchy patterns are seen for other meson nonets
and also in the baryon nonet and decuplet.

\subsection{The light scalar meson nonet}

\begin{figure}
\mbox{
  \includegraphics[height=0.4\textwidth,width=0.5\textwidth]{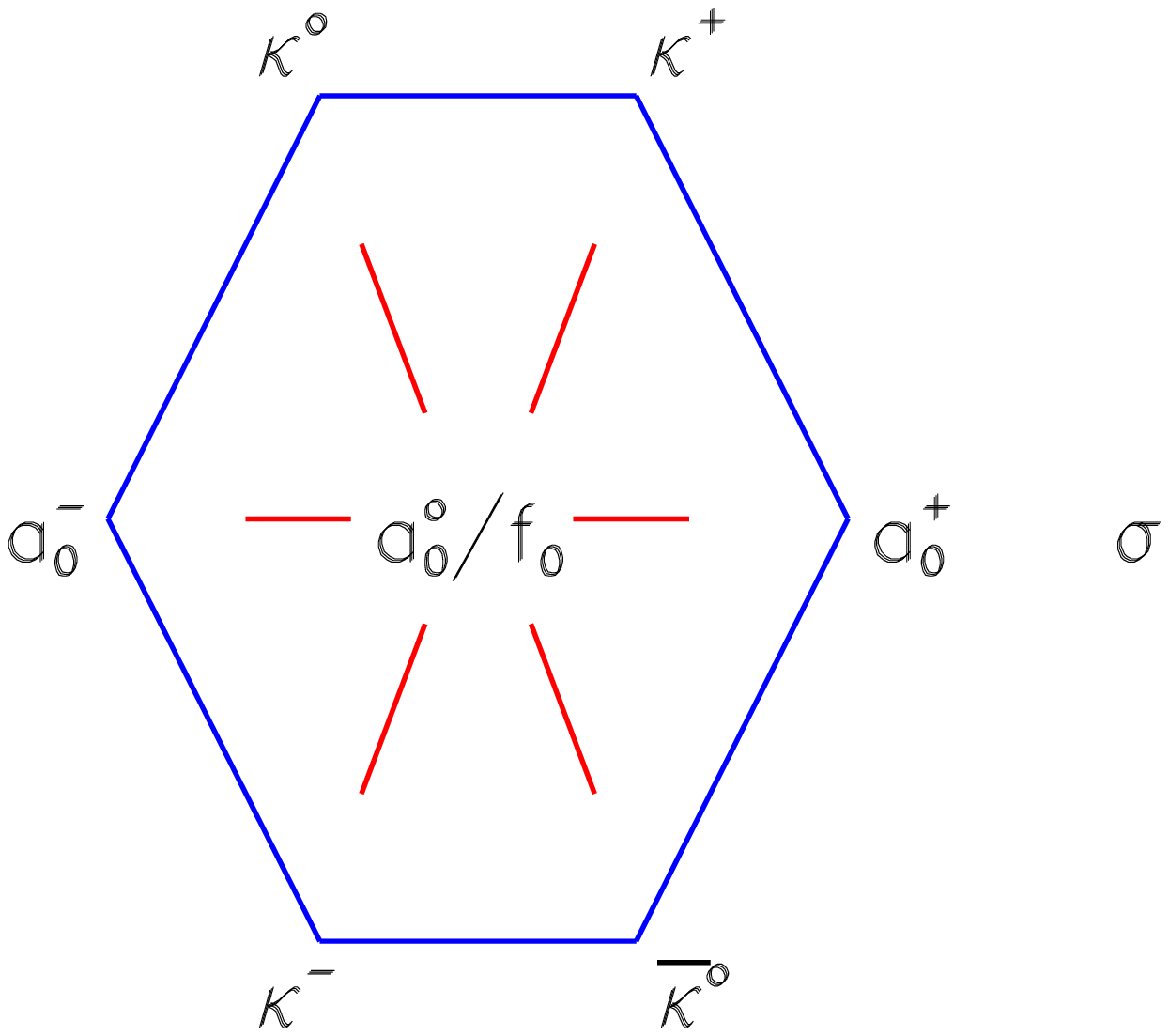}
}
\mbox{
  \includegraphics[height=0.4\textwidth,width=0.5\textwidth]{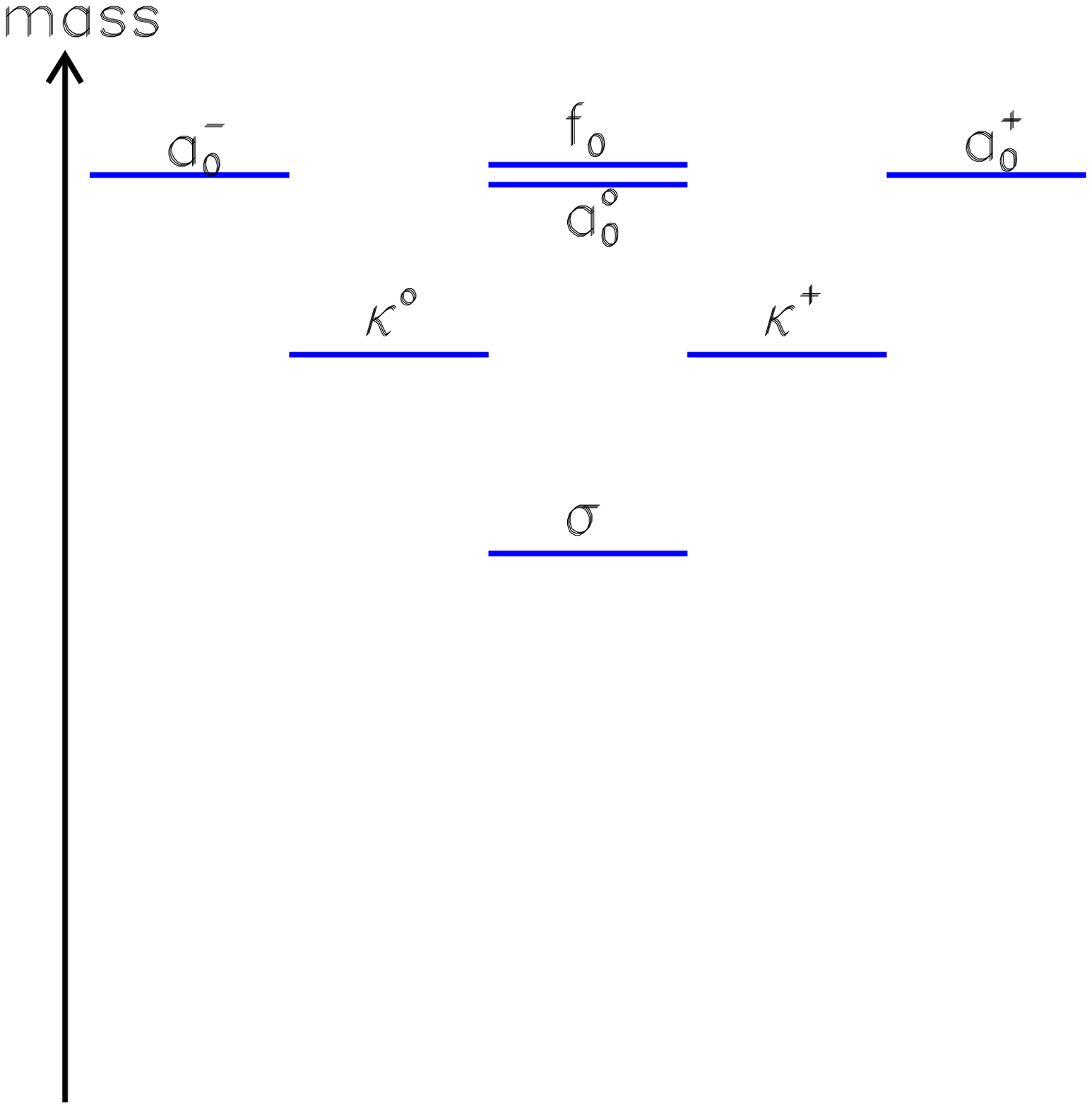}
}
\caption{{\bf Left:} the light scalar meson nonet.
{\bf Right:} the pseudoscalar
meson mass hierarchy.} 
\label{fig:scalar_nonet}
\end{figure}

In spite of the quark model's great acclaim, it has always had considerable
difficulty accounting for the light-mass scalar mesons, especially the
$a_0^{+,-,0}(980)$ isospin
triplet of mesons with $m\simeq 980$~MeV,\footnote{This report uses units
with $c=1$.}
 $\Gamma\simeq 100$~MeV and $J^P=0^+$,
seen in $\pi\eta$ spectra produced in two-photon collisions and $\ppbar$
annihilations~\cite{pdg},  and the $\fzero$, its isospin-singlet counterpart,
which has $J^{PC}=0^{++}$,
a similar mass \& width, and is seen via its $\fzero\rt \pipi$ decay mode in
many experiments.  If these are associated with the very broad, but now pretty
well established $\kappa^{+,0}$ and $\sigma^0$ resonances,\footnote{These are
referred to as the $K^*_0(800)$ and $f_0(600)$ in the PDG tables.} we can
identify what is called
the light scalar nonet shown in the left panel of Fig.~\ref{fig:scalar_nonet}.

For this nonet the meson mass hierarchy, shown in the right panel of 
Fig.~\ref{fig:scalar_nonet}, is inverted;  the $a_0(980)$ triplet, which in the
quark model does not have any $s$-quark content, has the highest mass.  In
addition, since the $\fzero$ is nearly degenerate with the  $\azero$, 
the $\fzero$-$\sigma$ mixing angle $\phi_S$ (the counterpart of
$\phi_P$ in Eq.~\ref{eq:eta-mix}) should be nearly zero and, thus, the $\fzero$
should have very small $s$-quark content.  Moreover, the higher mass tails of both
the $\fzero$ and the $\azero$ peaks are seen to have strong couplings to
$K\bar{K}$ final states:  a BESII study of $\fzero$ production in
$\jpsi\rt\phi \pipi$ and $\phi K^+K^-$ determines the ratio of
$K^+K^-$ to $\pipi$ couplings of 
$g^{f_0}_{KK}/g^{f_0}_{\pi\pi} = 4.2\pm0.3$~\cite{bes2_gfkk}
and an $\azero^{\pm}\rt K_LK^{\pm}$ signal seen in $\ppbar\rt K_L K^{\pm}\pi^{\mp}$ in the
Crystal Barrel experiment is used to infer
$g^{a_0}_{KK}/g^{a_0}_{\eta\pi} = 1.03\pm0.14$~\cite{xtal-bar_gakk}.  The
high masses and strong couplings to $K$ mesons suggest that the $a_0(980)$
and $f_0(980)$ have substantial $s$-quark content, which is not possible
in the simple $\qqbar$ picture for these mesons.

Another problem with a $\qqbar$ characterization of the light scalar-meson nonet
is that the $\qqbar$ pair has to be in a relative $P$-wave, in which case
there should be other $P$-wave nonets with $J^{PC} = 1^{+-}$, $1^{++}$ and $2^{++}$
nearby in mass.  The lowest mass nonets with these quantum numbers have masses
around 1300$\sim$1500~MeV and are generally associated with a well established
$0^{\, ++}$ nonet in this mass region.

If the $\azero$ and $\fzero$ are not $\qqbar$ states, what are they?  In a
classic 1977 paper, Jaffe suggested that these states might be tightly
bound tetraquark states~\cite{jaffe}.  This was based on the realization
that two colored quarks inside a hadron are most strongly attracted if they are
in a anti-triplet in color space.  This anti-color-triplet ``diquark'' can combine 
with two anticolored
antiquarks in a corresponding color-triplet combination  --a ``diantiquark''-- to form
a tightly bound color-singlet diquark-diantiquark mesonic state.  A different
possibility, suggested by Weinstein and Isgur and motivated by the proximity
of the $a_{\, 0}$ and $f_{\, 0}$ masses to the $K\bar{K}$ mass threshold,  
is that they are loosely bound 
molecule-like configurations of $K$ and $\bar{K}$ mesons~\cite{isgur}.

\subsection{$\azero$-$\fzero$ mixing?}

\begin{figure}
\begin{center}
\mbox{
  \includegraphics[height=0.35\textwidth,width=0.48\textwidth]{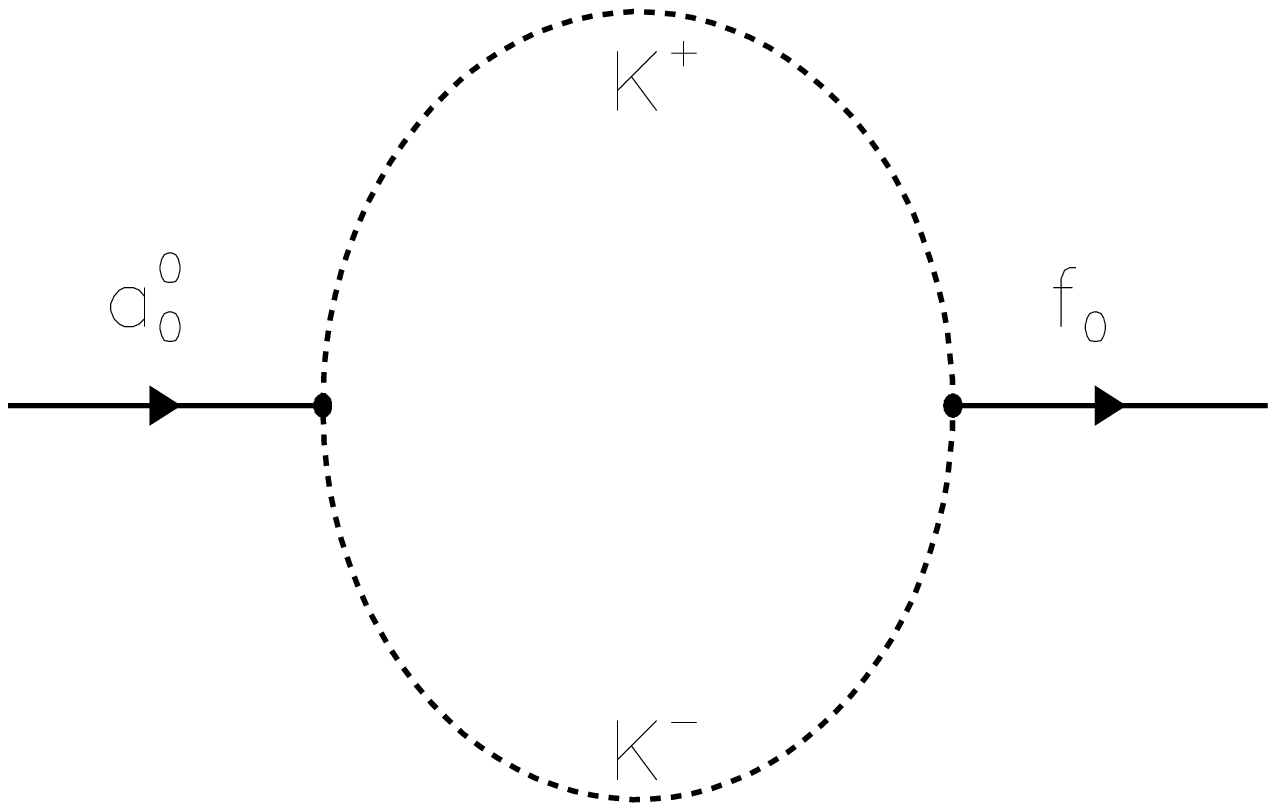}
}
\mbox{
  \includegraphics[height=0.35\textwidth,width=0.48\textwidth]{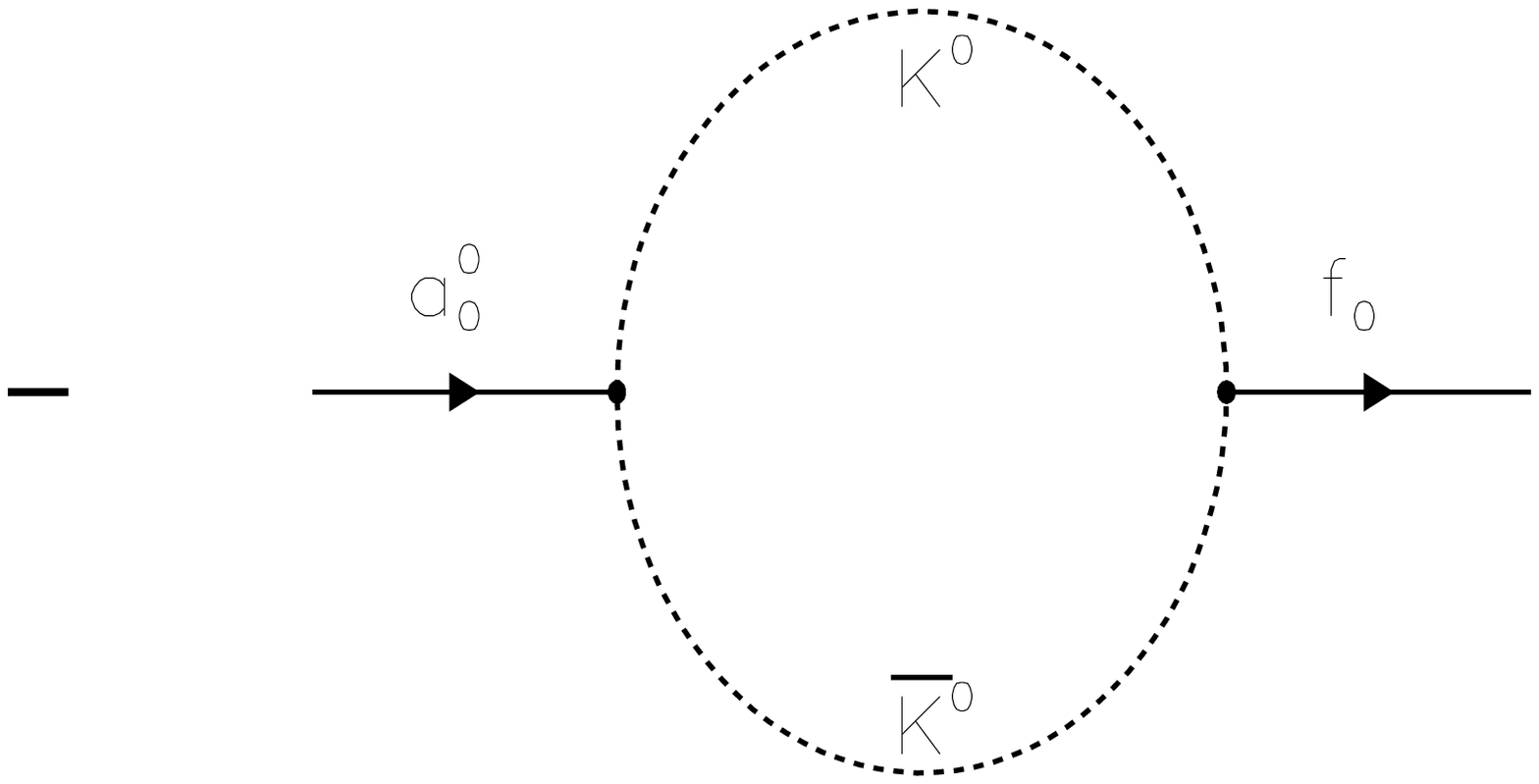}
}
\caption{The primary diagram for $\azero$-$\fzero$
mixing~\cite{hanhart}.  
} 
\label{fig:a0-f0-mix}
\end{center}
\end{figure}

Hanhart and collaborators~\cite{hanhart} have suggested that the
isopsin-violating $\azero\leftrightarrow\fzero$ process
would be a useful probe of the internal structure of the $\azero$-$\fzero$
meson system. 
The leading diagram for mixing of this type is the difference between
the $K^+K^-$ and $K^0\bar{K^0}$ loop processes shown
in Fig.~\ref{fig:a0-f0-mix}.  For $\azero)$ (or $\fzero$)
mass values that are either below $2m_{K^+}$ or above  $2m_{K^0}$,
the two diagrams more-or-less cancel each other out.  However,
for masses above $2m_{K^+}$ but below $2m_{K^0}$, the left-hand loop
is on the mass shell and is not strongly cancelled by the
right-hand loop, which is still off the mass shell.  Thus,
the mixing strength is strongly enhanced for the
$\sim$8~MeV mass window between the $K^+K^-$ and $K^0\bar{K^0}$
thresholds. 

\begin{figure}
\mbox{
  \includegraphics[height=0.4\textwidth,width=0.5\textwidth]{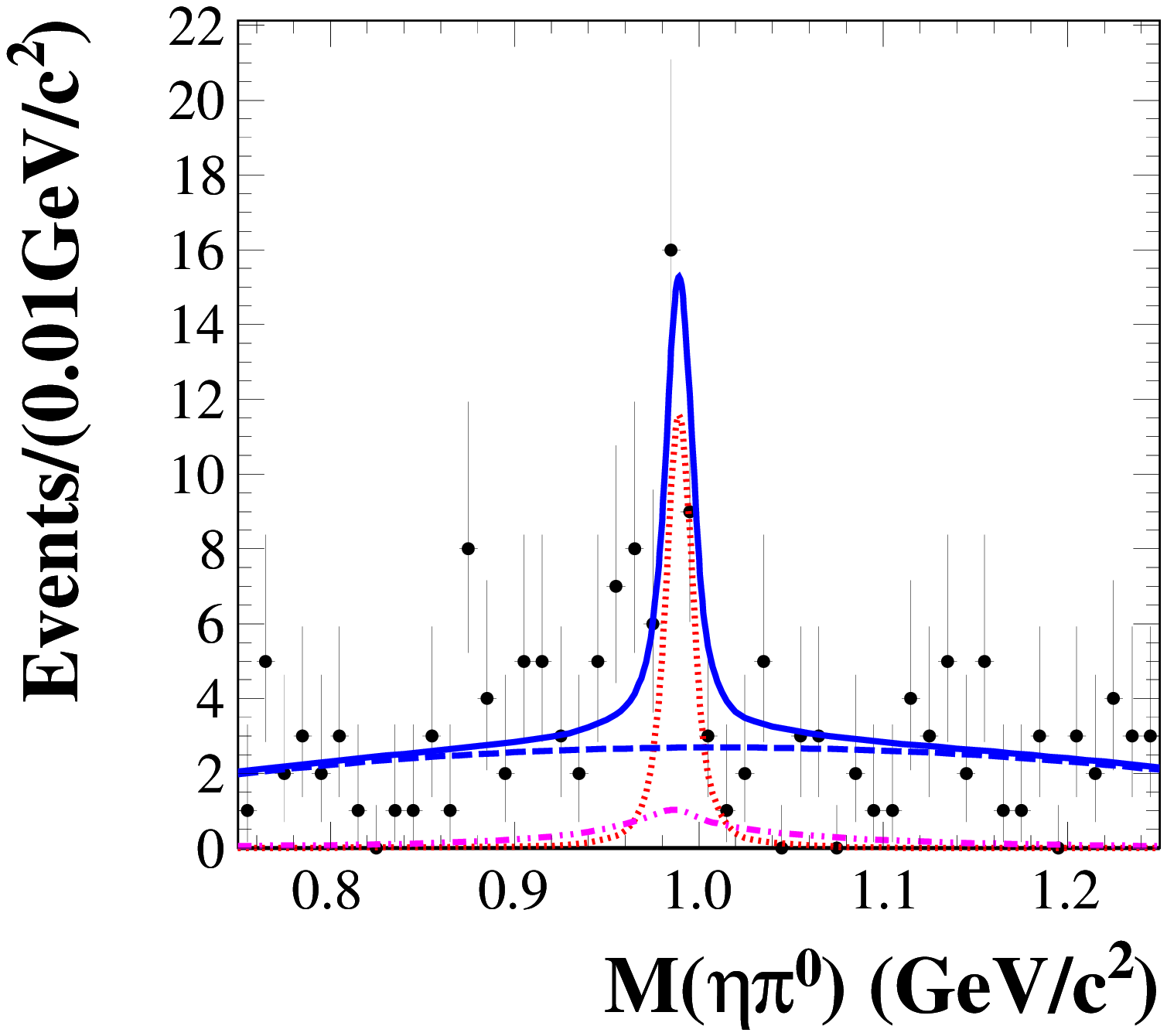}
}
\mbox{
  \includegraphics[height=0.4\textwidth,width=0.5\textwidth]{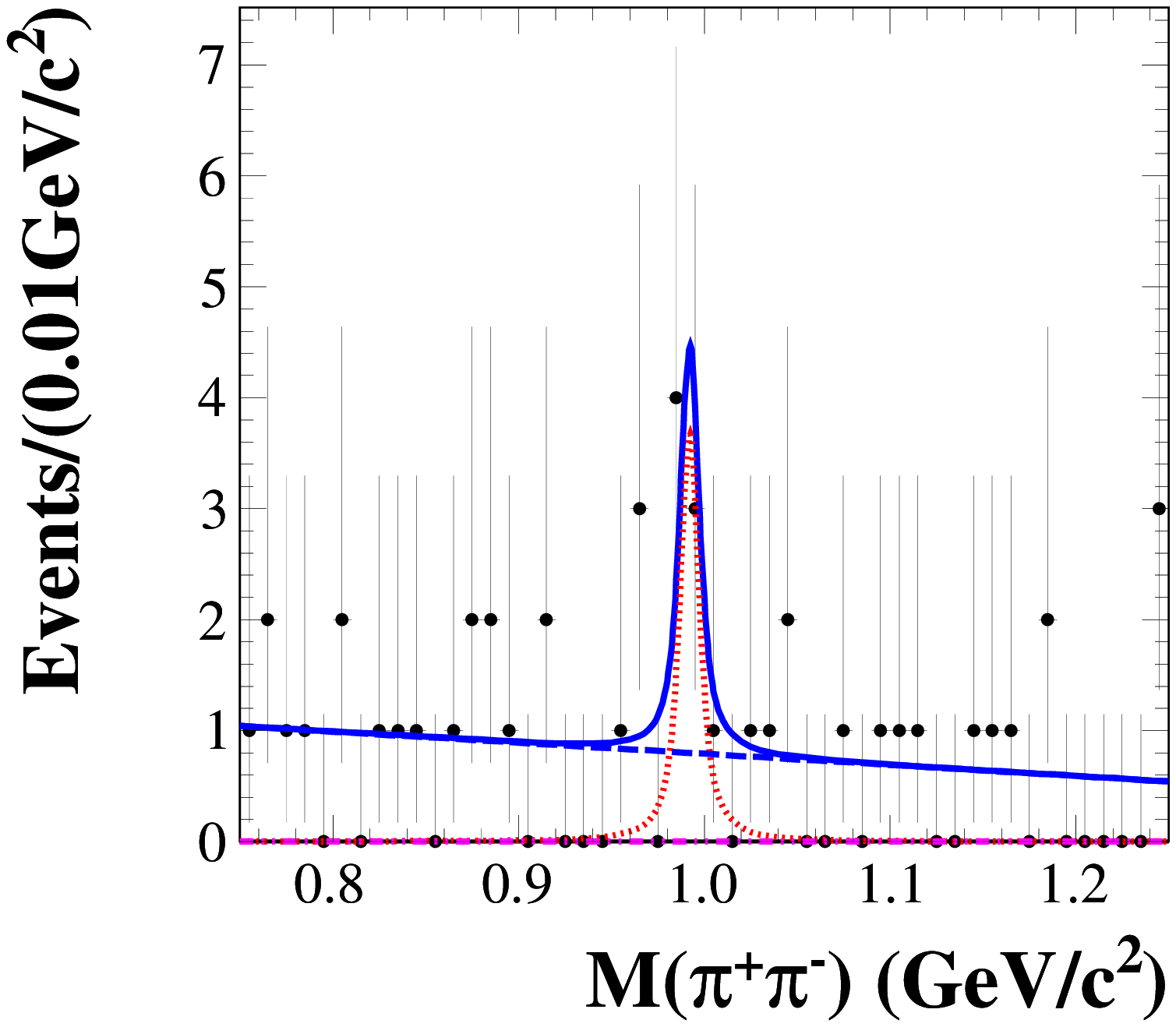}
}
\caption{{\bf Left:} The $\piz\eta$ mass distribution from $\jpsi\rt \phi \piz\eta$
decays.
{\bf Right:} The $\pipi$ mass distribution from 
$\psip\rt\gamma\chi_{c1}\rt\gamma\piz\pipi$ decays.} 
\label{fig:a0-f0-mix_data}
\end{figure}

BESIII searched for evidence for  $\fzero\rightarrow\azero$
mixing in the process $\jpsi\rt\phi f_0\rt\phi a^0_0\rt K^+K^-\eta\piz$
using the 225M event  $\jpsi$ data sample, and for evidence
for  $\azero\rightarrow\fzero$ mixing in the process
$\psip\rt\gamma\chi_{c1}\rt\gamma\piz a_0^0\rt\gamma\piz f_0\rt\gamma\piz\pipi$
using the 110M event $\psip$ data sample~\cite{bes3_a0mix}. 
The $\eta\piz$ invariant mass distribution from the $\jpsi\rt\phi a_{\, 0}$
search is shown in the left panel of Fig.~\ref{fig:a0-f0-mix_data},
where a prominent narrow peak is seen at the expected location.
The $\pipi$ mass distribution from the 
$\psi'\rt \gamma\chi_{c1}\rt \gamma\piz f_0$
search is shown in the right panel of Fig.~\ref{fig:a0-f0-mix_data}.
Here some excess of events at the expected location can be seen.

\begin{figure}
\begin{center}
\includegraphics[height=0.5\textwidth,width=0.5\textwidth]{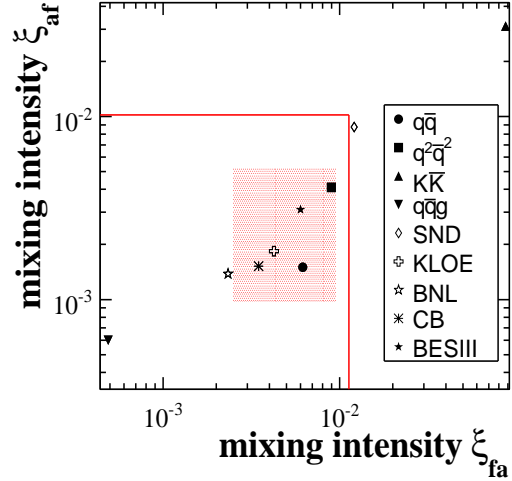}
\caption{The shaded square indicates the $\pm 1\sigma$ ranges
for the measured values of $\xi_{fa}$ and $\xi_{af}$ the 
$f_0 \rt a_0$ and $a_0\rt f_0$ mixing probabilities, respectively.
The solid red lines indicate 90\% CL upper limits.  The 
points represent expectations for different models and 
line-shape parameterizations (see Ref.~\cite{bes3_a0mix}). } 
\label{fig:a0-f0_limits}
\end{center}
\end{figure}

From the fits shown as solid curves in the figures, BESIII determines
probabilities for $f_{\, 0}$ mixing to $a^0_{\, 0}$, $\xi_{fa}=(0.60\pm 0.34)$\%
or $<1.1$\% at the 90\% confidence level (CL),
and for $a_{\, 0}$ mixing to $f_{\, 0}$, $\xi_{af}=(0.31\pm 0.21)$\% 
or $<1.0$\% at the 90\% CL.\footnote{Here, and elsewhere in this report, 
statistical, systematic
and model-dependence errors are added in quadrature.} 

These limits imply that ($\xi_{fa},\xi_{af}$) values outside 
of the box indicated by the solid red lines in Fig.~\ref{fig:a0-f0_limits}
are ruled out;  this includes the values expected for a pure
$K\bar{K}$ molecule picture that are indicated in the
figure by the solid triangular point near $(\xi_{fa},\xi_{af})=(0.1,0.03).$

\subsection{Isospin violations in $\eta(1405)$ decays}

\begin{figure}
\mbox{
  \includegraphics[height=0.4\textwidth,width=0.5\textwidth]{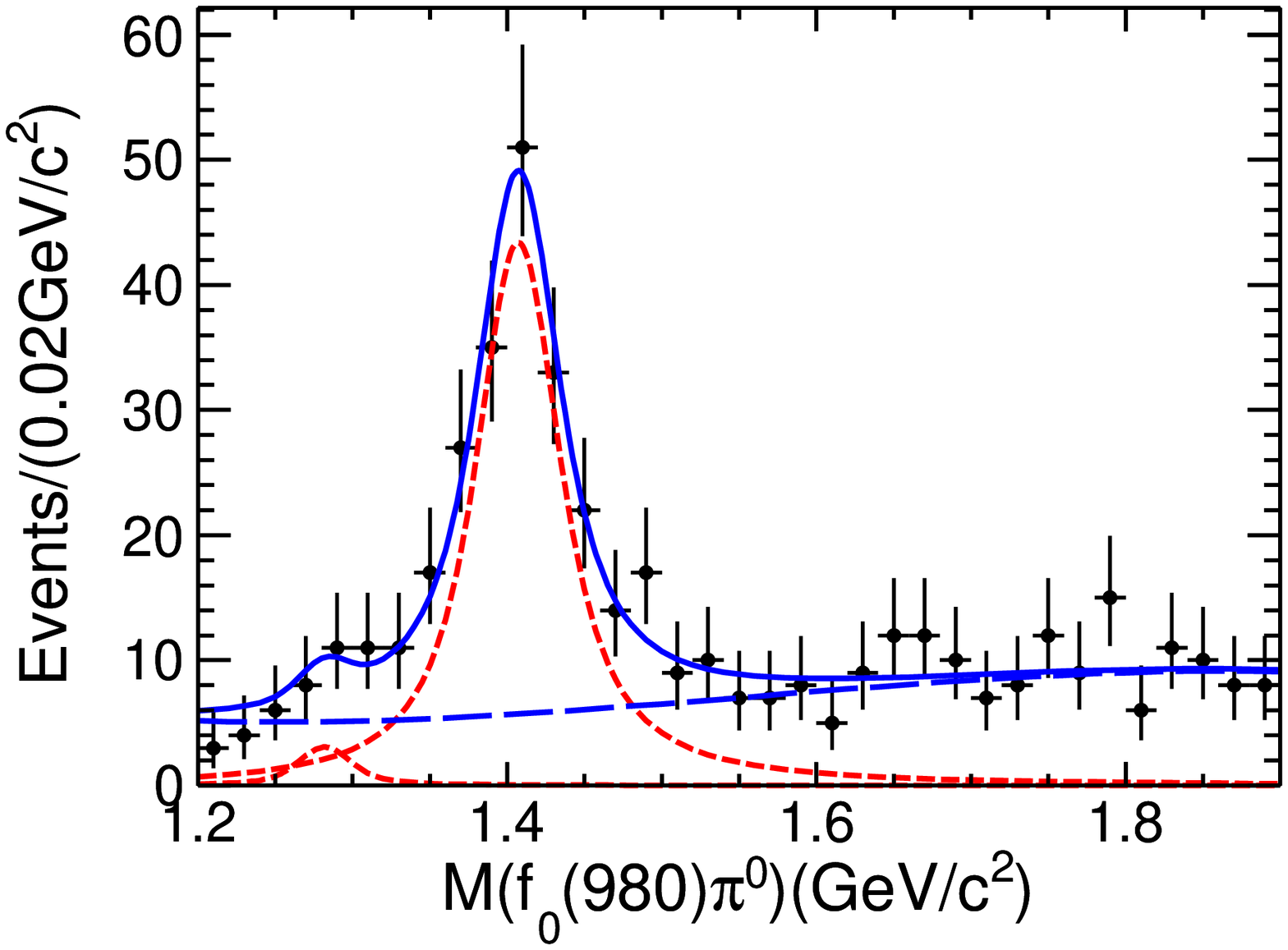}
}
\mbox{
  \includegraphics[height=0.4\textwidth,width=0.5\textwidth]{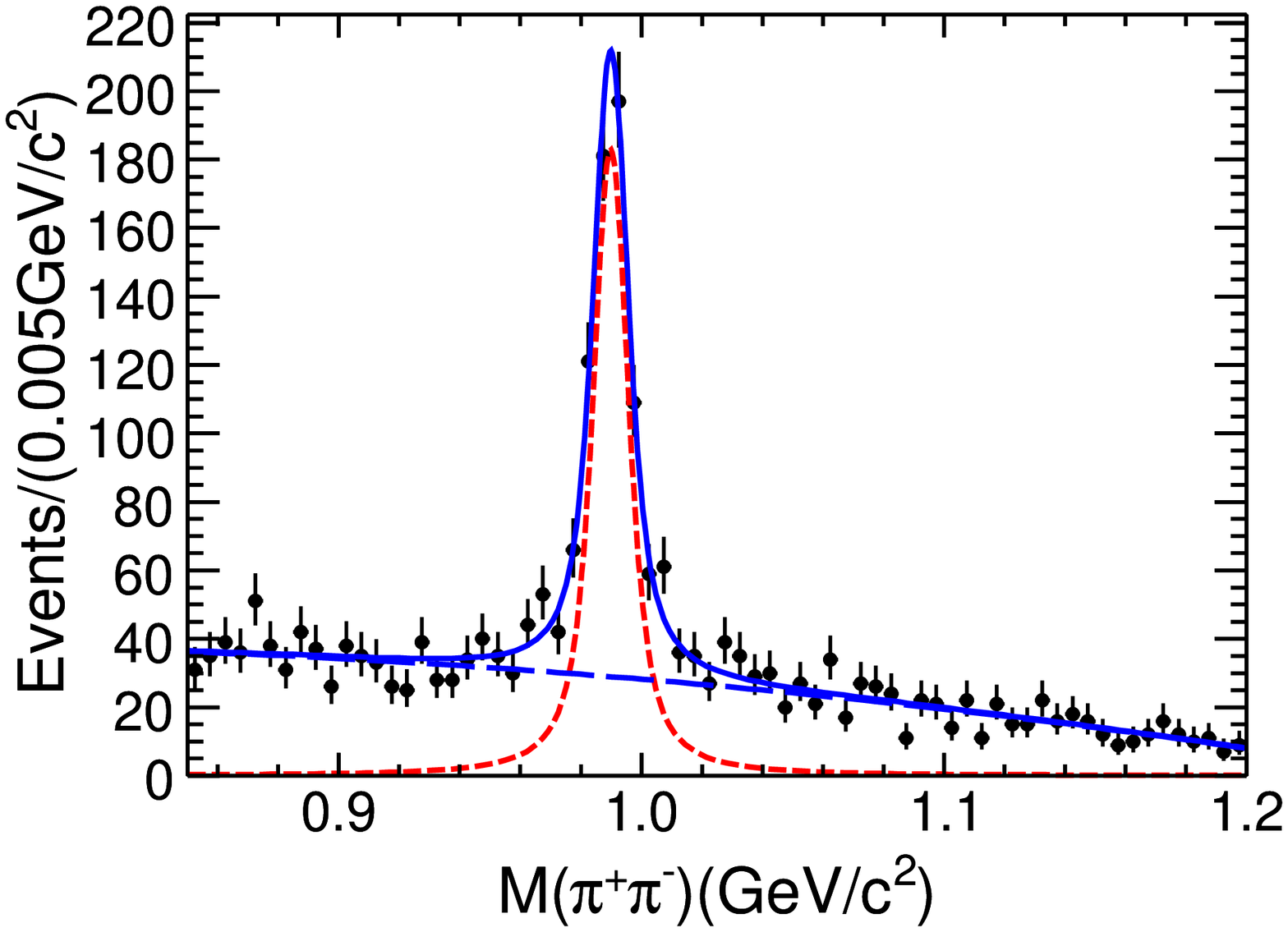}
}
\caption{{\bf Left:} The $\piz f_0$ mass distribution from $\jpsi\rt \gamma \piz f_0$,
decays where $f_0\rt\piz\piz$ (a $\jpsi\rt 7\gamma$ final state!).
{\bf Right:} The $\pipi$ mass distribution from 
$\eta(1405)\rt\piz f_0$ decays where $f_0\rt\pipi$.} 
\label{fig:pi0f0}
\end{figure}

In a separate study, BESIII examined the $\piz f_0$ invariant
mass distribution produced in radiative $\jpsi\rt \gamma\piz f_0$
decays for both the $f_0\rt \pipi$ and $f_0\rt\piz\piz$ decay
modes~\cite{bes3_eta1405}.  In the distribution for $f_0\rt \piz\piz$ decays, shown in
the left panel of Fig.~\ref{fig:pi0f0} (the $f_{\,0}\rt\pipi$ channel
looks similar), the dominant feature is a pronounced
peak near $M(\piz f_0)=1405$~MeV; helicity analyses
indicate that this peak has $J^p=0^-$, which leads to its
identification as the $\eta(1405)$ resonance.

The decay $\eta(1405)\rt\piz f_0$ violates isospin.  In this
case the observed isospin violation is quite large:
\begin{equation}
\frac{Bf(\eta(1405)\rt\piz \fzero\rt\piz\pipi)}{Bf(\eta(1405)\rt \piz\azero\rt\piz\piz\eta)}
= (17.9\pm 4.2)\%,
\end{equation}
which is an order-of-magnitude larger than is typical for isospin violations. (For
example, BESIII also reports that the isospin violating
$Bf(\etap\rt\pipi\piz)$ is ($0.9 \pm 0.1$)\% of the isospin
conserving $Bf(\etap\rt\pipi\eta)$~\cite{bes3_eta1405}.)

A striking feature 
of these decays is the lineshape of the $f_{\, 0}\rt\pi\pi$ decays,
shown for the $f_{\, 0}\rt\pipi$ channel in the right panel of
Fig.~\ref{fig:pi0f0}, where it can be seen that the
$f_{\, 0}$ peak position is significantly above its nominal $980$~MeV value, and
its width is much narrower than its nominal value of $\sim$100~MeV.
The fitted mass is $M=989.9\pm0.4$~MeV, midway between 
$2m_{K^+}$ and $2m_{K^0}$, and the fitted width is $\Gamma = 9.5\pm 1.1$~MeV,
consistent with the $2m_{K^0} - 2m_{K^+} = 7.8$~MeV mass threshold
difference.

\begin{figure}
\begin{center}
\includegraphics[height=0.6\textwidth,width=0.7\textwidth]{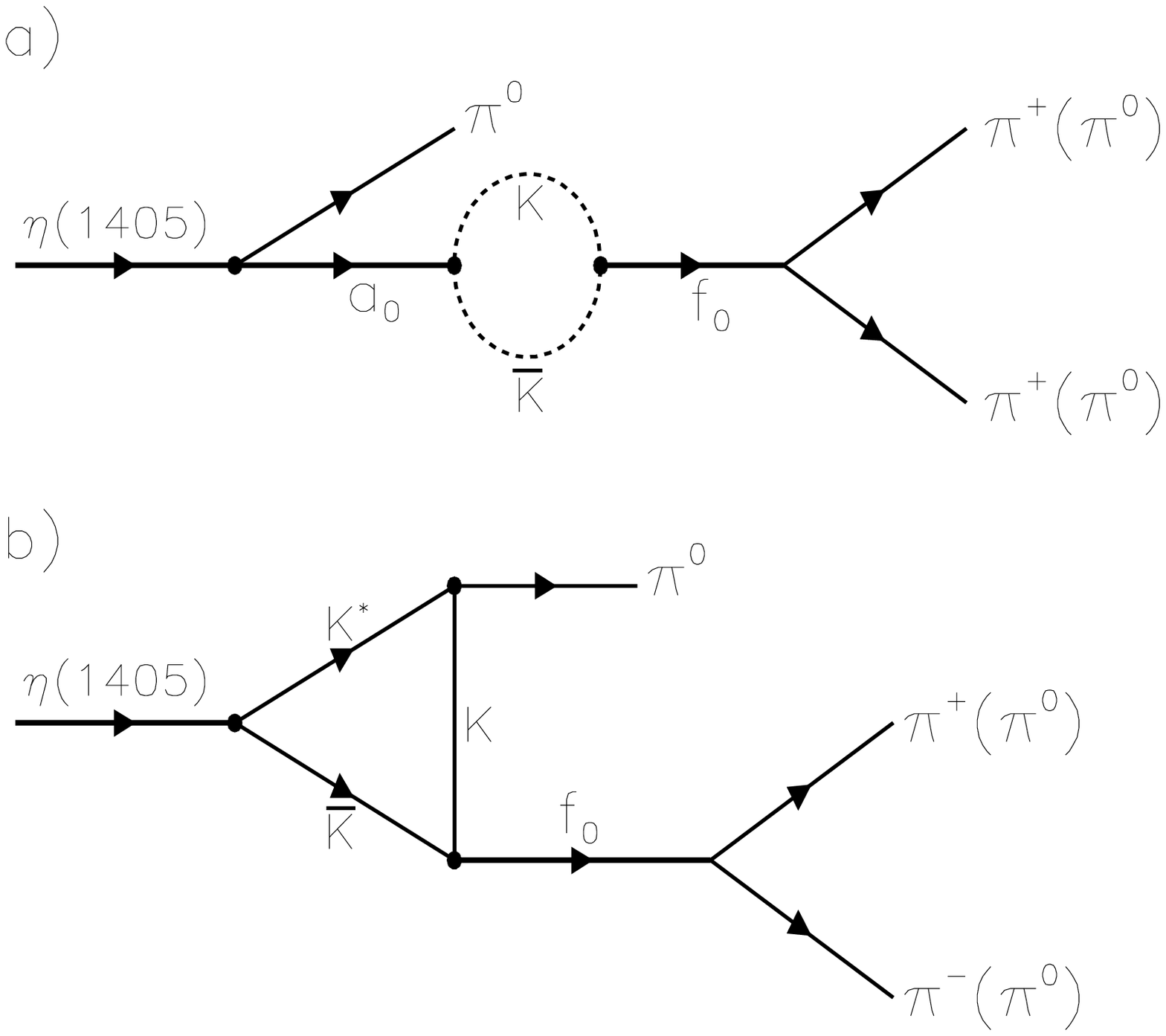}
\caption{{\bf a)} The leading diagram for $\eta(1405)\rt\piz f_0$
via $\azero$-$\fzero$ mixing.  
{\bf b)} The triangle anomaly
diagram in $\eta(1405)\rt \piz\fzero$ decay~\cite{zhao}.
} 
\label{fig:eta1405_triangle}
\end{center}
\end{figure}

Possible processes that mediate $\eta(1405)\rt\piz f_0$ are shown in
Fig.~\ref{fig:eta1405_triangle}.  As we have seen above, the 
$\azero\rt\fzero$ process (Fig.~\ref{fig:eta1405_triangle}a)
is at or below the percent level, and
is too small to account for the large isospin violation that
is observed.  Wu and collaborators~\cite{zhao} suggest that the 
triangle anomaly diagram shown in Fig.~\ref{fig:eta1405_triangle}b
could be large enough to account for the data.  In this case, both
the $K^*\bar{K}$ system that couples to the $\eta(1405)$ and the
$K\bar{K}$ system coupling to the $f_{\, 0}$ can have large on-mass-shell,
isospin-violating contributions.

While our understanding of the low mass scalar mesons remains unclear, 
it seems that detailed studies -- both theoretical and experimental --
of isospin violations in processes involving the $a_{\; 0}(980)$ and $f_0(980)$ 
can provide important probes of their inner workings.  The results
presented above are from data samples that are small fractions of
what we ultimately expect to collect with BESIII.  With the full
data sets we will be able to provide theorists with  precision measurements
of the $a_0(980)\leftrightarrow f_0(980)$ mixing parameters and other quantities
related to these mesons.

\section{Some puzzles in $\jpsi$ and $\psip$ decays}

The charmonium mesons are (nearly) non-relativistic bound $\ccbar$ pairs and
probably the simplest hadron system to understand.  However, a number of
puzzling features have been observed that seem to defy conventional understanding.
I briefly mention a few of these here.

\subsection{$\psip\rt\gamma\eta$ puzzle}

\begin{figure}
\begin{center}
\includegraphics[height=0.7\textwidth,width=0.7\textwidth]{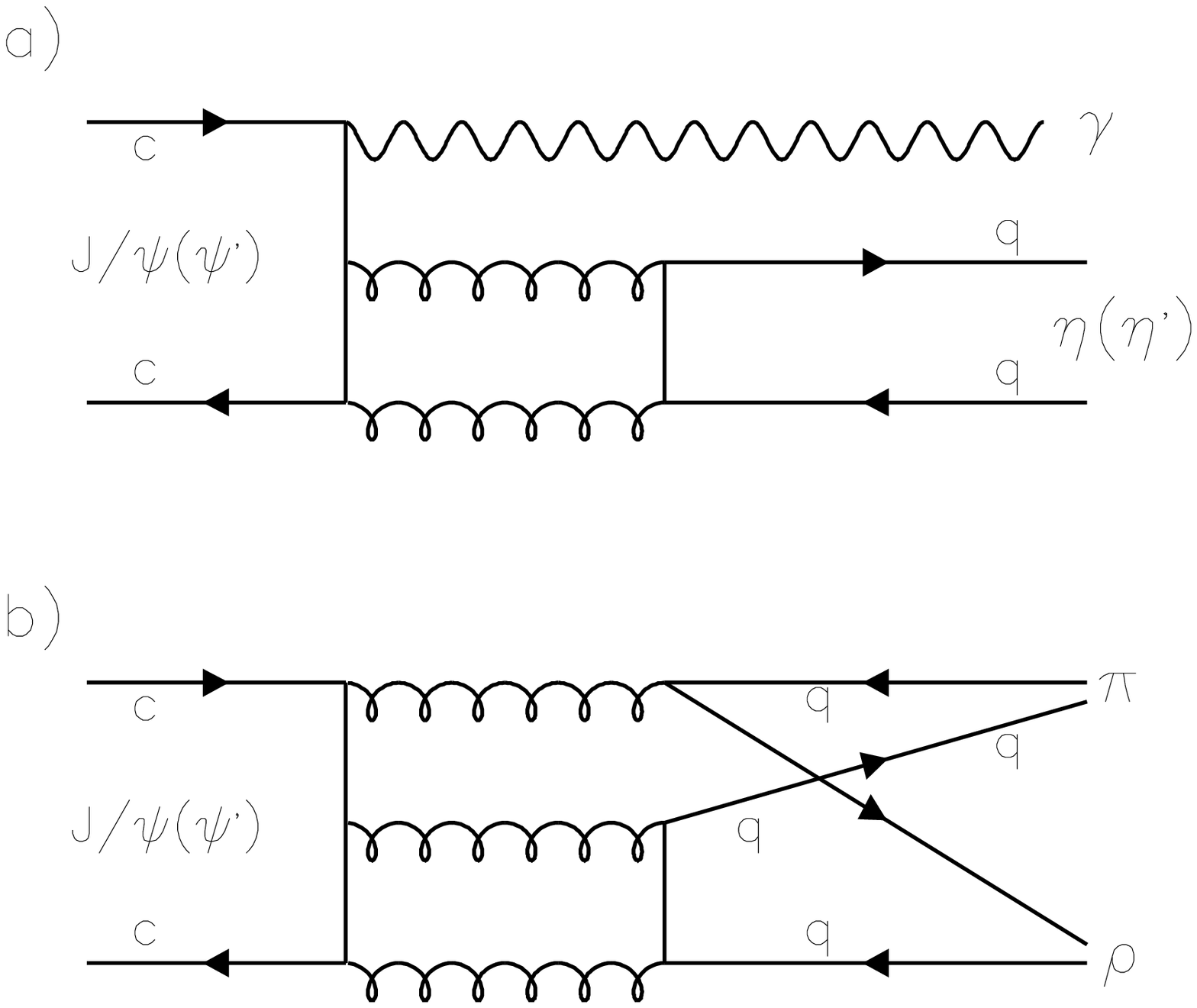}
\caption{The lowest-order quark-line diagrams for {\bf a)} $\jpsi\ (\psip)\rt\gamma\eta\ (\etap)$
and {\bf b)} $\jpsi\ (\psip)\rt\rho\pi$.  
} 
\label{fig:jpsi_decays_2box}
\end{center}
\end{figure}

The expected lowest-order diagram for
$\jpsi \rt \gamma \eta$ and $\jpsi\rt \gamma \etap$ is
illustrated in Fig.~\ref{fig:jpsi_decays_2box}a.  If this
diagram dominates, the ratio of partial widths for the
two processes can be related to the $\eta$-$\etap$ mixing
angle via the relation
\begin{equation}
\frac{\Gamma(\jpsi\rt\gamma\eta)}{\Gamma(\jpsi\rt\gamma\etap)}
=(\frac{p_{\etap}}{p_{\eta}})^3\frac{1}{\tan^2\theta_P},
\label{eq:gam-eta}
\end{equation}
where $\theta_P$ is the pseudoscalar mixing angle in the
$SU(3)$ basis\footnote{This related to angle $\phi_P$ used
in Eq.~\ref{eq:eta-mix} by $\theta_P =\phi_P - \arctan\sqrt{2}=\phi_P-54.7^{\circ}$.} 
and $(p_{\etap}/p_{\eta})^3$ is a phase-space factor~\cite{cahn}. From the
PDG values for branching fractions, the ratio of
partial widths is $0.21\pm0.01$, which gives $|\theta_{~P}| = (22.5\pm 0.5)^{\circ} $,
consistent with the value determined from other methods.

The same diagram is expected to apply to $\psip\rt\gamma\eta$ and $\gamma\etap$
decays in which case the same relation to the mixing angle should apply.
However, although the mode $\psip\rt\gamma\etap$ was well established
long ago~\cite{bes2_gam-etap}, only just recently did BESIII report~\cite{bes3_gam-eta}
the first observation of a signal for $\psip\rt\gamma\eta$.
For some unknown reason, the $\psip\rt\gamma\eta$ mode is strongly suppressed:
the partial width ratio for the $\psip$ is only $0.011\pm0.004$, more than an order-of-magnitude
lower than the corresponding ratio for the $\jpsi$.  If this partial-width ratio
is used in Eq.~\ref{eq:gam-eta}, the resulting value for $|\theta_{~P}|$
is $(5.6\pm 0.9)^{\circ}$, much smaller than the value determined from $\jpsi$ decays.

\subsection{The $\rho\pi$ puzzle: a new twist to an old story}

The oldest puzzle in charmonium physics is the so-called $\rho\pi$ puzzle.
$\jpsi\rt\rho\pi$ is the strongest hadronic decay mode of the $\jpsi$, with
a branching fraction of $(1.69\pm 0.15)$\%~\cite{pdg}.  The lowest-order
diagram for this decay is expected to be the three-gluon annihilation process
shown in Fig.~\ref{fig:jpsi_decays_2box}b.   The same diagram is expected to
apply to the $\psip$ and, thus, the partial width $\Gamma(\psip\rt\rho\pi)$
is expected to be that for the $\jpsi$, scaled by the ratio of the
$\ccbar$ wavefunctions at the origin and a phase-space factor.  (The ratio
of the wavefunctions at the origin is determined by comparing the
$\jpsi\rt\ee$ and $\psip\rt\ee$ partial widths.)  The result of this reasoning
is the famous ``12\% rule,'' which says that the branching fraction
for $\psip$ to some hadronic state should be (roughly) 12\% that of
the $\jpsi$ to the same final state.  While this simple rule more-or-less
works for many decay modes, it fails miserably for $\psip\rt\rho\pi$
decays, where $Bf(\psip\rt\rho\pi)=(3.2\pm 1.2)\times10^{-5}$, nearly a factor
of a hundred below the 12\%-rule expectation.

\begin{figure}
\mbox{
  \includegraphics[height=0.4\textwidth,width=0.5\textwidth]{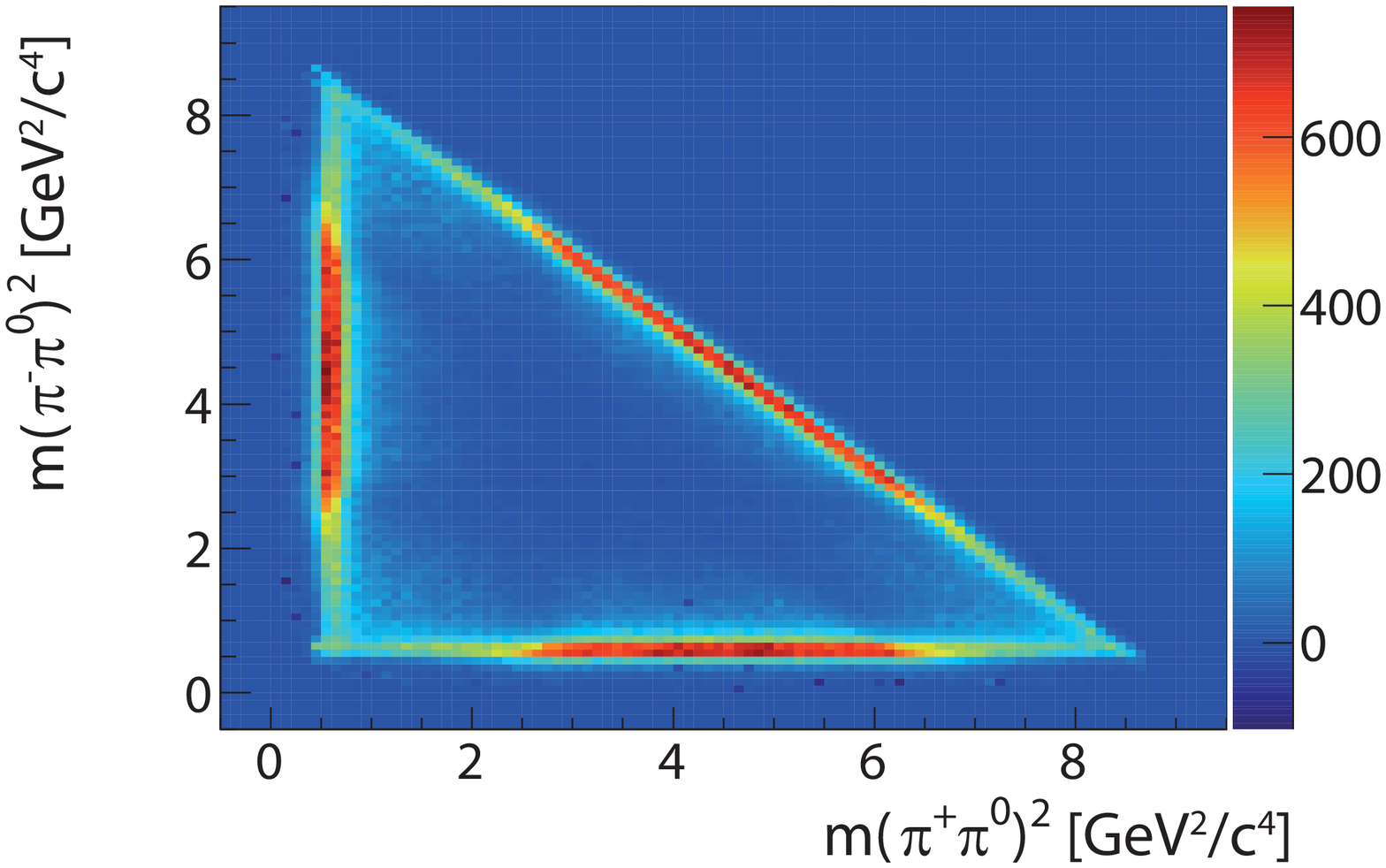}
}
\mbox{
  \includegraphics[height=0.4\textwidth,width=0.5\textwidth]{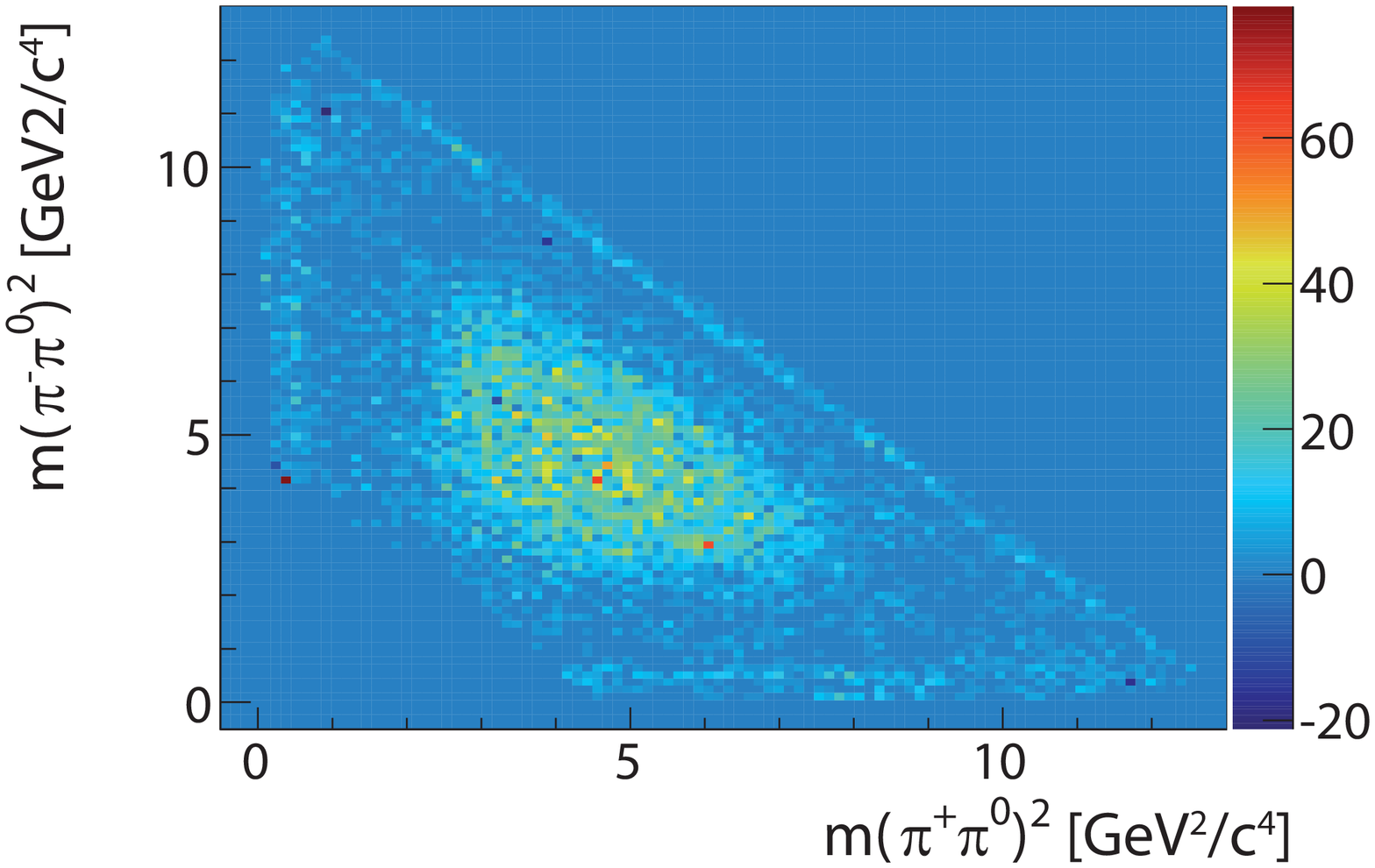}
}
\mbox{
  \includegraphics[height=0.3\textwidth,width=0.5\textwidth]{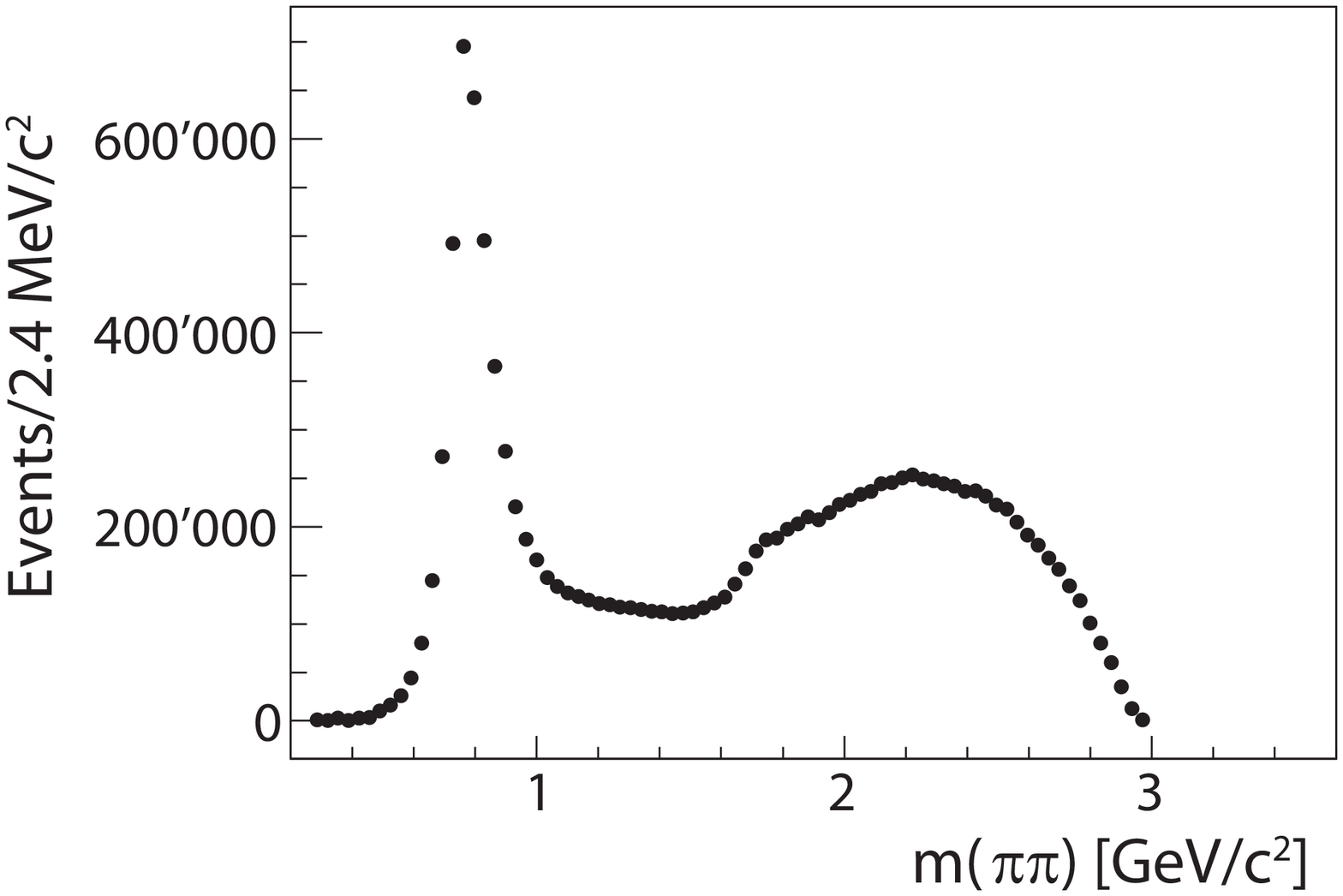}
}
\mbox{
  \includegraphics[height=0.3\textwidth,width=0.5\textwidth]{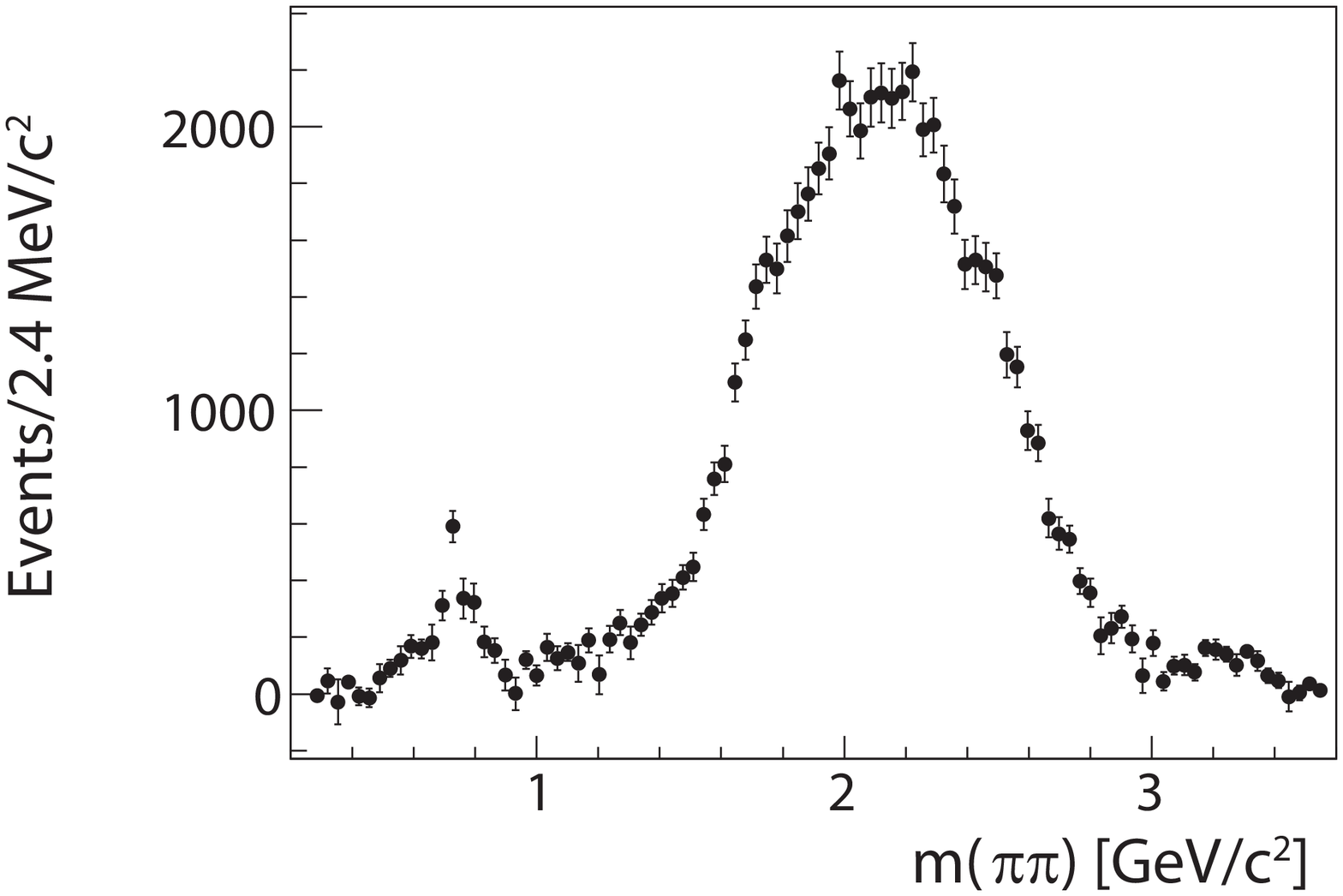}
}
\caption{{\bf Top:} the $M^2(\pi^-\piz)$ (vertical) {\it vs.} $M^2(\pi^+\piz)$
(horizontal) for (left) $\jpsi\rt\pipi\piz$ and 
  (right) $\psip\rt \pipi\piz$ decays.
{\bf Bottom:} the $M(\pi\pi)$ projections of the Dalitz plots. 
}
\label{fig:rhopi}
\end{figure}

BESIII has recently reported on a high-statistics study of of
$\jpsi\rt\pipi\piz$ and $\psip\rt\pipi\piz$~\cite{bes3_rhopi} using the
225M event $\jpsi$ and 106M event $\psip$ data samples. 
The $M^2(\pi^-\piz)$ (vertical) 
{\it vs.} $M^2(\pi^+\piz)$ (horizontal) Dalitz plot
distributions, shown in the top panels of Fig.~\ref{fig:rhopi}, 
for the $\jpsi$ (left) and $\psip$ (right) data samples, could not be more
different.  The center of the $\jpsi\rt\pipi\piz$ Dalitz plot
is completely devoid of events, while in the $\psip\rt\pipi\piz$
plot most of the events are concentrated in the center.  The dynamics
of the two processes are completely different, in spite of the fact that
the underlying process --illustrated in Fig.~\ref{fig:jpsi_decays_2box}b--
is expected to be very similar.  The $\rho\pi$ puzzle is becoming even more
puzzling.

\subsection{The subthreshold $\ppbar$ resonance seen in $\jpsi \rt \gamma \ppbar$}

As mentioned above in the introduction, BESII reported a peculiar
mass-threshold enhancement in the $\ppbar$ invariant mass distribution
in radiative $\jpsi\rt\gamma\ppbar$ decays~\cite{bes_x1860}.  The
shape of this low-mass peak cannot be reproduced by any of the 
commonly used parameterizations for final state interactions (FSI)
between the final-state $p$ and $\bar{p}$. 

\begin{figure}
\begin{center}
\includegraphics[height=0.35\textwidth,width=0.7\textwidth]{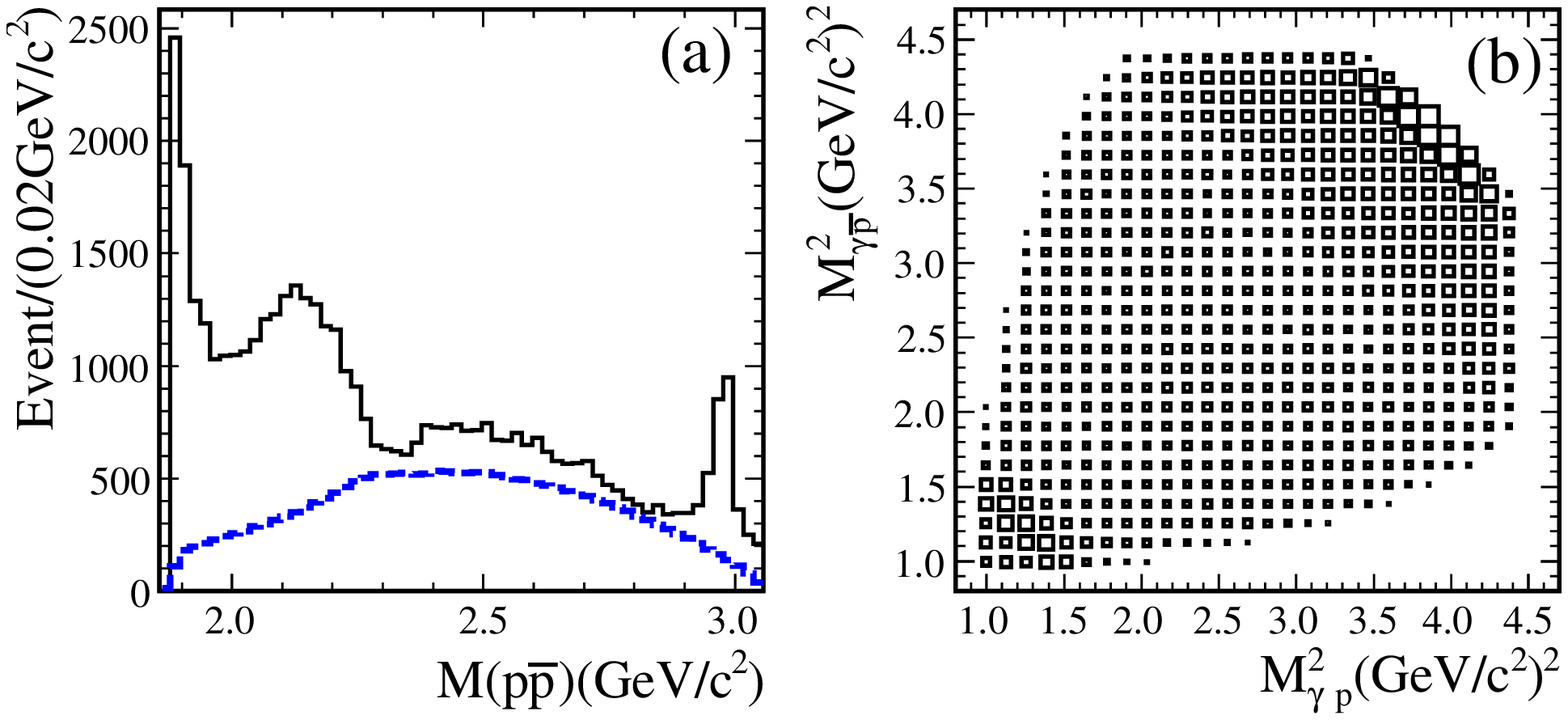}
\caption{{\bf a)} The $M(\ppbar)$ distribution from $\jpsi\rt\gamma\ppbar$ 
decays.  The dashed curve is background from $\jpsi\rt\piz\ppbar$,
where one of the photons from the $\piz\rt\gamma\gamma$ decay has low energy
and is undetected.  The narrow peak on the right is from 
$\jpsi\rt\gamma\etac$, $\etac\rt\ppbar$.
{\bf b)}  The $M^2(\gamma\bar{p})$ (vertical) {\it vs.} $M^2(\gamma p)$
Dalitz plot for the same data sample.  The diagonal band at the upper right
is produced by the $\ppbar$ mass-threshold enhancement; the band at the
lower left is due to the $\etac$.
}
\label{fig:x1860}
\end{center}
\end{figure}

The $\ppbar$ invariant mass
distribution for $\jpsi\rt\gamma\ppbar$ decays in the 225M event BESIII $\jpsi$ data sample
is shown in Fig.~\ref{fig:x1860}a, where the threshold enhancement
is quite prominent~\cite{bes3_x1860}.   A Dalitz plot for these events
is shown in Fig.~\ref{fig:x1860}b.  A partial-wave-analysis (PWA)
applied to these data determined that the $J^{PC}$ of the near-threshold
structure is $0^{-+}$.   A fit using a sub-threshold
resonance shape modified by the Julich FSI effects~\cite{julich_fsi}
yields a mass of $M=1832^{+32}_{-26}$~MeV
and a 90\% CL upper limit on the width of $\Gamma<79$~MeV.

One intriguing theoretical speculation about this state is that it is a
nucleon-antinucleon bound state, sometimes called baryonium.  In this context
some authors suggested that it may also decay to $\pipi\etap$ at a 
substantial rate~\cite{yan}.  A BESII search for a corresponding state in
$\jpsi\rt\gamma\pipi\etap$ decays found a previously unseen resonance,
dubbed the $X(1835)$, with peak-mass value of $1834\pm 7$~MeV and with a width
of $\Gamma = 68\pm 22$~MeV~\cite{bes2_x1835}, both of which are in good
correspondence with the fitted parameters of the $\ppbar$ subthreshold
resonance peak and in accord with the prediction of Ref.~\cite{yan}.

\begin{figure}
\mbox{
  \includegraphics[height=0.4\textwidth,width=0.5\textwidth]{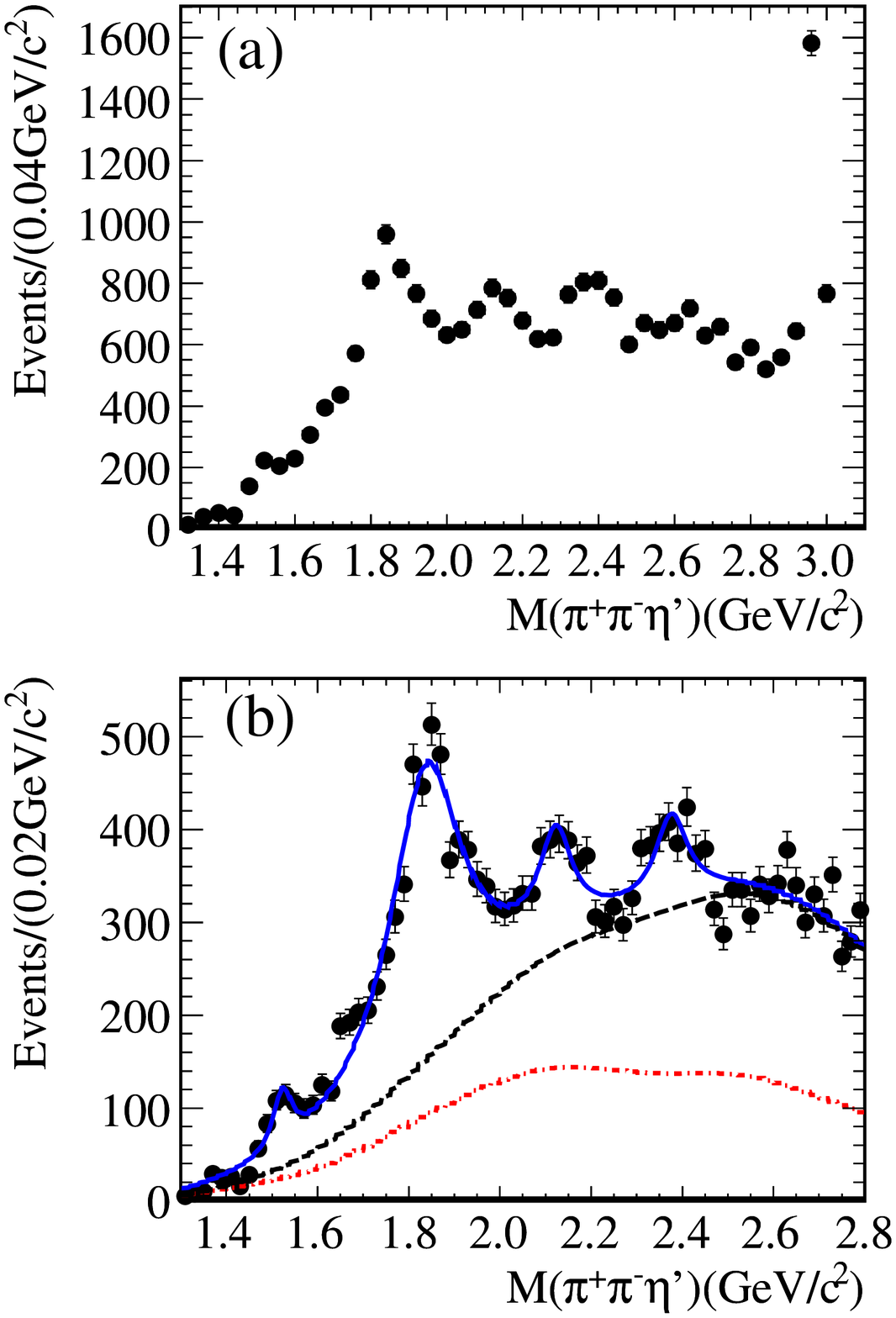}
}
\mbox{
  \includegraphics[height=0.3\textwidth,width=0.5\textwidth]{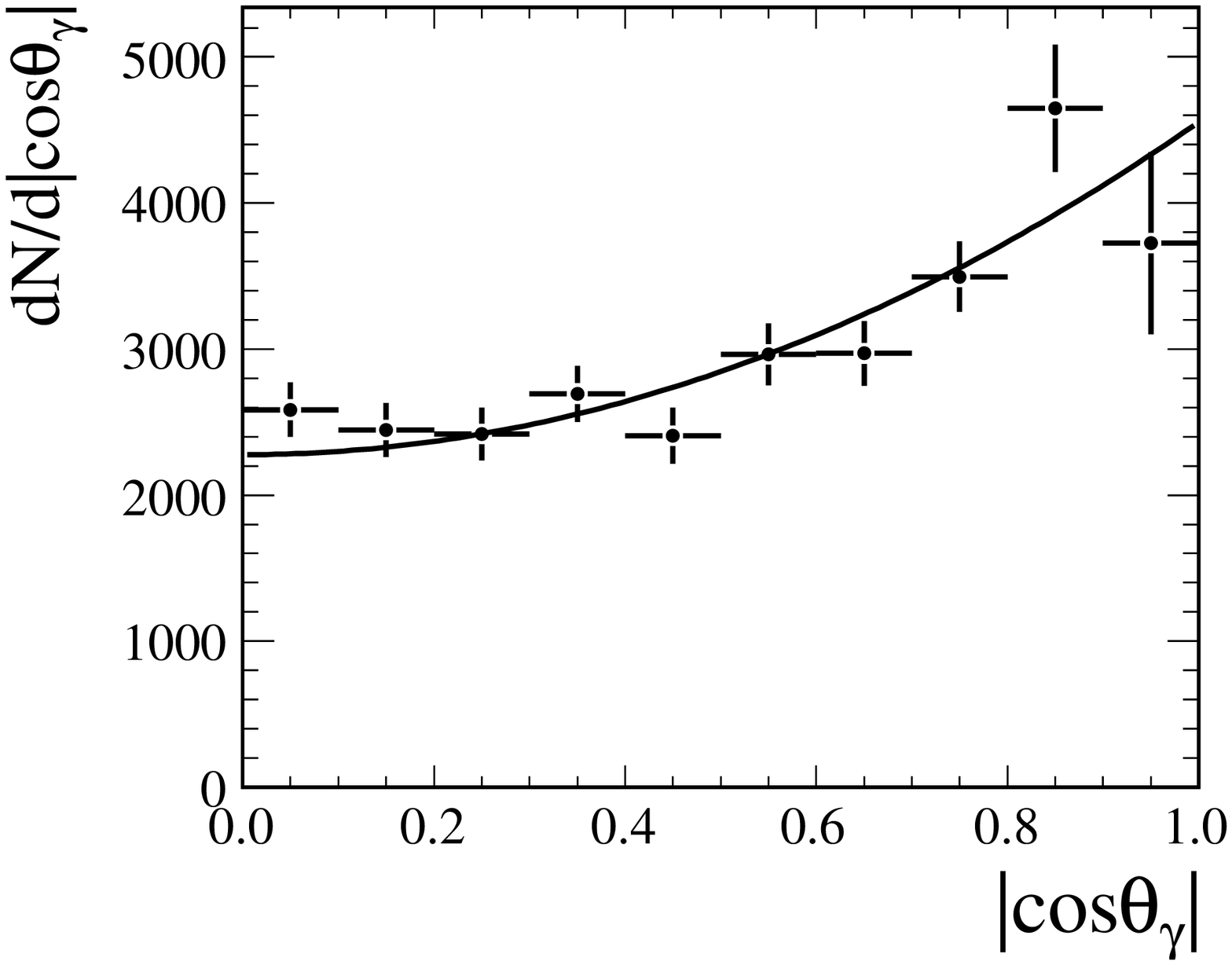}
}
\caption{{\bf Left:} The $\pipi\etap$  
mass distribution from $\psip \rt \gamma\pipi\etap$.  The
red dash-dot curve in the bottom panel indicates the
non-$\etap$ background determined from the $\etap$ mass side-bands.
{\bf Right:} The $X(1835)$ event yield in bins of $|\cos\theta_{\gamma}|$,
where $\theta_{\gamma}$ is the polar angle of the angle of the 
radiated photon.} 
\label{fig:x1835}
\end{figure}

Recently reported BESIII measurements of the $\pipi\etap$ mass spectrum
in $\jpsi\rt\gamma\pipi\etap$ are shown in the left panel of
Fig.~\ref{fig:x1835}, where a peak corresponding to the $X(1835)$
is evident~\cite{bes3_x1835}.  While the fitted value 
for the peak mass, $M=1837^{+6}_{-4}$~MeV, is consistent with the
BESII results for both the $\pipi\etap$ and $\ppbar$ final states,
the new results for the width are much larger, $\Gamma= 190\pm 39$~MeV.
The angular distribution of the radiated $\gamma$, shown in the right 
panel of Fig.~\ref{fig:x1835}, is consistent the $1+ \cos^2\theta_{\gamma}$ 
form expected for $J^{PC}=0^{-+}$.  
Although the mass and $J^{PC}$ values are consistent with those in 
the $\ppbar$ channel, the broad width is not.  However, the width
determination of the $\ppbar$ peak has complications because of
its sub-threshold character --we only see a small part of its high mass tail-- 
while the $X(1835)\rt\pipi\etap$ peak has a large 
underlying background of real $\pipi\etap$ events
that likely has components that interfere
with the $X(1835)$ resonance amplitude, a possibility that is
not considered in the Ref.~\cite{bes3_x1835} fit for the $X(1835)$
mass and width.

An unexpected feature in the Fig.~\ref{fig:x1835} (left) spectrum is the
existence of two additional, rather pronounced peaks at higher masses. Fitted
values for the masses and widths of these peaks are
\begin{eqnarray}
M_1 &=& 2122^{+8}_{-7}~{\rm MeV};~~~\Gamma_1=83^{+35}_{-19}~{\rm MeV}\\ 
M_2 &=& 2376^{~+9}_{-10}~{\rm MeV};~~~\Gamma_2=83^{+47}_{-18}~{\rm MeV}.
\end{eqnarray}
A striking characteristic of these peaks is their relatively narrow widths.  Ordinary
light-hadron resonances that are so far above threshold are expected
to be very wide; there is no previously established light-hadron meson resonance
with a width this narrow and a mass above 2~GeV.  An intriguing
possibility is that these states may be the $2^{++}$ and $0^{-+}$ 
glueballs for which an unquenched lattice QCD calculation
predicts  masses of $2390\pm 125$~MeV and $2560\pm 125$~MeV,
respectively~\cite{chen_lqcd}.  With the
huge additional $\jpsi$ event sample expected from future BESIII
running, PWA will be used to determine the $J^{PC}$ values of these peaks,
which may help clarify their underlying nature.

\section{Precise measurements of properties of the $\etac$ and $h_c$ charmonium states}

\begin{figure}
\begin{center}
\includegraphics[height=0.7\textwidth,width=0.7\textwidth]{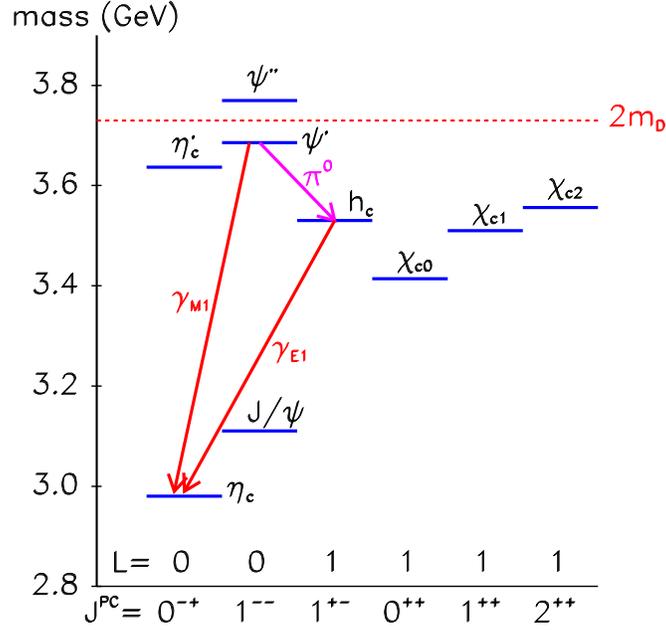}
\caption{The spectrum of the low-lying charmonium mesons.  The red dashed
line indicates the $M=2m_{ D}$ open-charmed threshold. States with mass above
this value can decay to final states containing $D$ and $\bar{D}$ mesons
and are typically broad; states below this threshold are relatively narrow.
The magenta and red arrows indicate transitions used for the $\etac$
and $h_c$ measurements reported here.   
} 
\label{fig:charmonium}
\end{center}
\end{figure}

The charmonium mesons are important because of their simplicity and their accessibility
by a variety of theoretical approaches, including effective field theories and
lattice QCD~\cite{nora}.
Because of their large mass, the charmed quarks bound in the charmonium meson states
have relatively low velocities, $v^2\sim 0.3$, and non-relativistic potential models
can be used with relativisitic effects treated as small perturbations.
With the discovery of the $\etac^{\prime}$ by Belle in 2002~\cite{belle_etacp} and the $h_c$
by CLEO in 2005~\cite{cleo_hc}, all of the charmonium states below the $M=2m_{~D}$ open-charm
threshold have been identified (see Fig.~\ref{fig:charmonium}).  
An experimental task now is the provision of precision
measurements that can challenge the various theories that address this system.  In
this report I discuss recent BESIII measurements of properties the $\etac$ and $h_c$
charmonium states.

\subsection{Meaurement of the $\etac$ mass and width}

The $\etac$ is the ground state of the charmonium system.  The
mass difference between the $\jpsi$ and the $\etac$ is 
due to hyperfine spin-spin interactions and is, therefore,
a quantity of fundamental interest.  However, while the mass
of the $\jpsi$ is known to very high precision --better than
4 PPM-- the $\etac$ mass remains poorly measured, the 2010
PDG world average (WA) value is $m_{\etac}^{~2010}= 2980.3 \pm 1.2$~MeV,
and the measurements that go into this average have poor
internal consistency: the CL of the fit to a single mass
is only 0.0018.  The $\jpsi$-$\etac$ hyperfine mass splitting
derived from this WA is $\delta_{hfs}= 116.6\pm 1.2$~MeV, a
value that has always been above theoretical predictions~\cite{bali}.
The $\etac$ width is also very poorly known; the 2010 PDG WA for
this, $\Gamma_{\etac}^{~2010}=28.6\pm 2.2$~MeV, has a confidence level of
only 0.0001.   

Measurements of the $\etac$ mass and width roughly fall into
two categories, depending on how the $\etac$ mesons used in the
measurement are produced.   Experiments using $\etac$ mesons 
produced via $\jpsi$ radiative transitions tend to find a low
mass ($\sim 2978$~MeV) and narrow width ($\sim 10$~MeV),
while measurements using $\etac$ mesons produced via two-photon
collisions or $B$-meson decays find higher mass and width values.
A primary early goal of the BESIII experiment has been to try
to clear up this situation.

\begin{figure}
\mbox{
  \includegraphics[height=0.4\textwidth,width=0.5\textwidth]{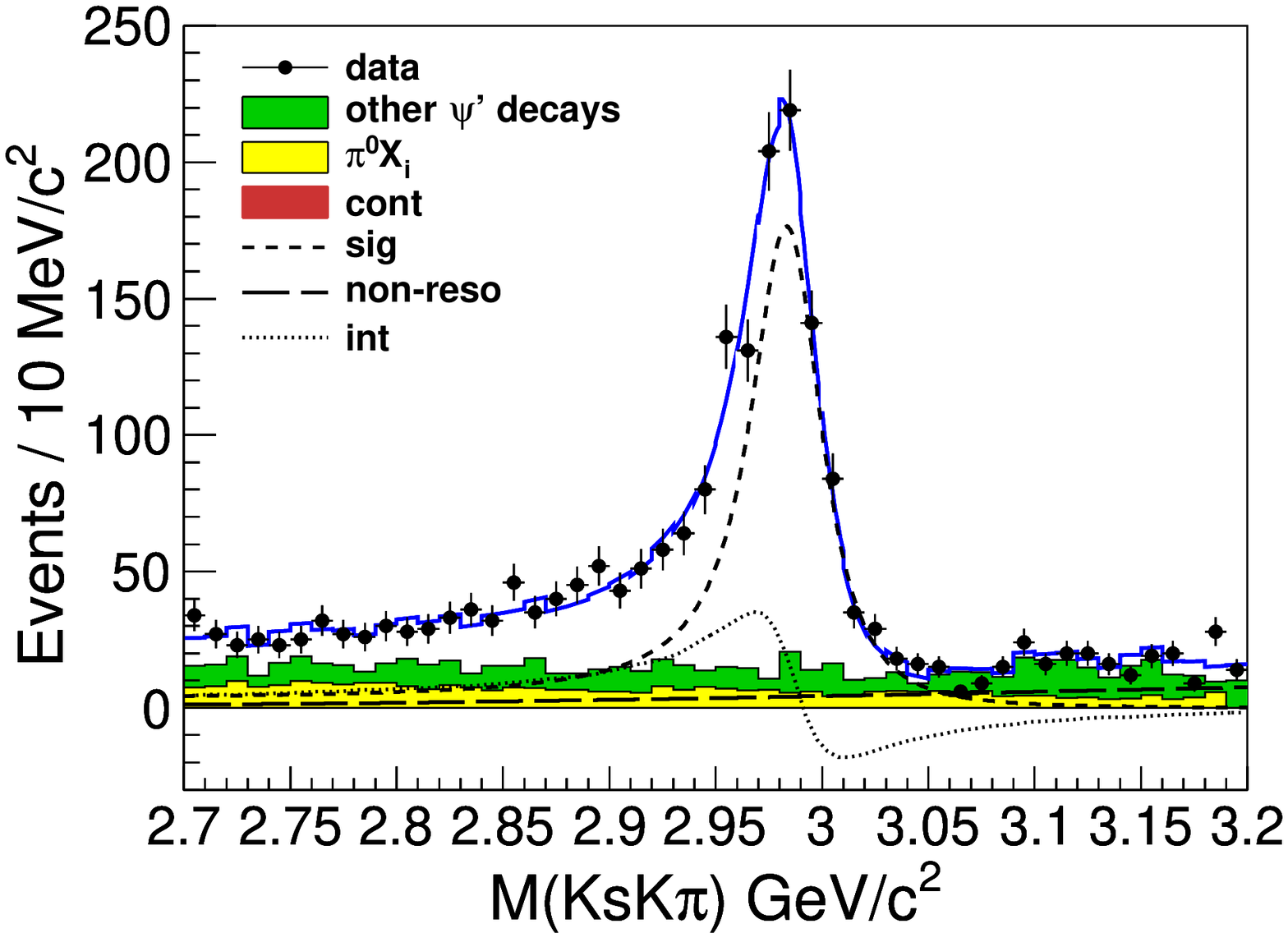}
}
\mbox{
  \includegraphics[height=0.4\textwidth,width=0.5\textwidth]{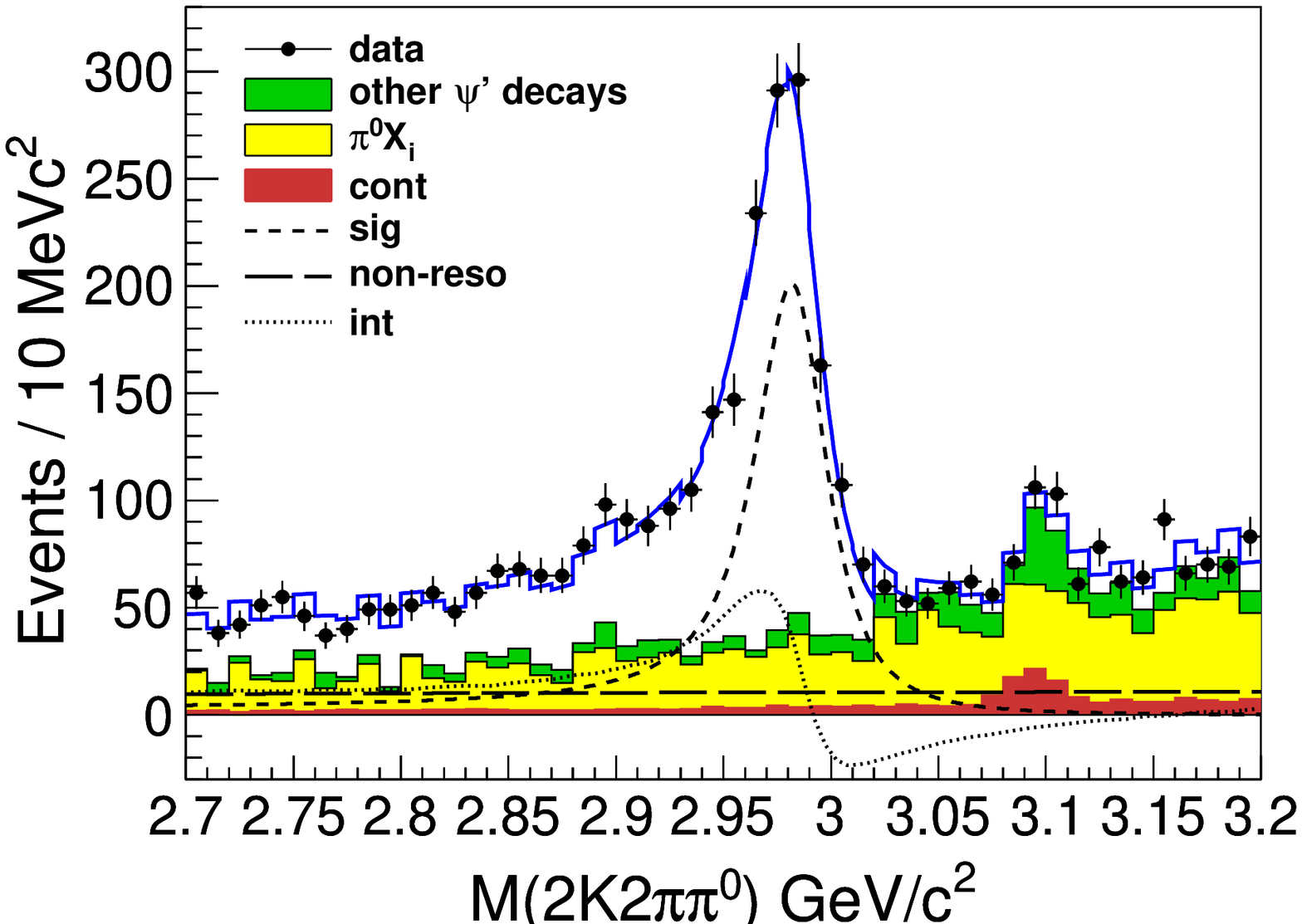}
}
\caption{{\bf Left:} The $K_SK^+\pi^-$ mass spectrum from  
$\psip \rt \gamma K_SK^+\pi-$ decays.
{\bf Right:} The corresponding plot for the $K^+K^-\pipi\piz$ channel.
The main background in most channels, indicated as yellow histograms, are from
$\psip\rt\piz X_i$, where $X_i$ is the same final state 
as the $\etac$ decay mode that is under study, and the $\piz\rt\gamma\gamma$
decay is asymmetric where one $\gamma$ has very low energy and
is not detected. This background is incoherent and does not interfere with
the $\etac$ signal.} 
\label{fig:etac_mass}
\end{figure}

A recently reported BESIII mass and width measurement~\cite{bes3_etac} uses
samples of $\etac$ mesons produced via the M1 radiative transition
$\psip\rt\gamma\etac$ (indicated by a red arrow in
in Fig.~\ref{fig:charmonium}) that decay to one
of six fully reconstructed final states:\footnote{
The inclusion of charge conjugate states is implied.}
$\etac\rt X_i$, where $X_i=K_SK^+\pi^-$, $K^+K^-\piz$, $\eta\pipi$,
$K_S K^+\pipi\pi^-$, $K^+K^-\pipi\piz$, and $3(\pipi)$,
where $K_S\rt\pipi$ and $\piz~(\eta)\rt\gamma\gamma$.   Distinct
$\etac$ signals are seen in each of the six channels, two
typical mass spectra are shown in Fig.~\ref{fig:etac_mass}.

In all six channels, the $\etac$ signal has a distinctively asymmetric
shape with a long tail at low masses and a rapid drop on the high mass
side.  This is suggestive of possible interference with a coherent
non-resonant background.   The solid blue curves in Fig.~\ref{fig:charmonium}
show the results of a fit that uses a Breit Wigner (BW) amplitude to
represent the $\etac$ that is weighted by a factor of $E_{\gamma}^7$ that
accounts for the M1 transition matrix element ($E_{\gamma}^3$)
and the wave-function mismatch between the radially excited $\psip$
and the ground-state $\etac$ ($E_{\gamma}^4$); the fit also allows for
interference with background from nonresonant $\psip\rt\gamma X_i$
decays.  Since fits to individual channels give consistent results for
the mass, width and the same value for the interference phase, a
global fit to all six channels at once with a single mass, width
and phase is used to determine the final results:
\begin{eqnarray}
m_{\etac} &=& 2984.3 \pm 0.8~{\rm MeV}\\
\Gamma_{\etac} &=& 32.0 \pm  1.6~{\rm MeV}.
\end{eqnarray}
The value of the phase $\phi$ depends upon whether the constructive or
destructive interference solution is used:  $\phi_{cons}=2.40\pm 0.11$
or $\phi_{des}=4.19\pm 0.09$. (The mass and width values for the
two cases are identical.)  The reason that the interference phase
is the same for all six channels is not understood.

The new BESIII mass and width values agree well with the earlier higher
values found in two-photon and $B$-meson decay meaurements.  The
probable reasons for the low values found by earlier measurements
using $\etac$ mesons produced via radiative charmonium decays are
the effects of the wave-function mismatch~\cite{cleo_etac}
and interference with the
non-resonant background that were not considered.  Using only the new
BESIII $\etac$ mass value, the $\jpsi$-$\etac$ hyperfine mass splitting
becomes smaller:
$\delta_{hfs}=112.6\pm 0.8$~MeV, and in better agreement with theory. 

\subsection{Meaurements of $h_c$ mass, width and branching fractions}

The singlet $P$-wave $h_c$ meson is notoriously difficult to study.  In
fact, despite considerable experimental efforts, it evaded detection
for some thirty years until it was finally seen by CLEO in 2005 in the
isospin-violating $\psip\rt\piz h_c$ transition (indicated by a
magenta arrow in Fig.~\ref{fig:charmonium})~\cite{cleo_hc}.
To date, it has only been seen by two groups, CLEO and
BESIII~\cite{bes3_hc1} and only via the 
strongly suppressed $\psip\rt\piz h_c$ process.

In lowest-order perturbation theory, the $h_c$ mass
is equal to the spin-weighted-average of the
triplet $P$-wave $\chi_{c0,1,2}$ states: 
$<m_{\chi_{cJ}}>=(m_{\chi_{c0}}+3m_{\chi_{c1}}+5m_{\chi_{c2}})/9=3525.30\pm 0.04$~MeV.
Theoretical predictions for the  branching fraction for $\psip\rt\piz h_c$
are in the range $(0.4\sim 1.3)\times 10^{-3}$, the
E1 radiative transition $h_c\rt \gamma\etac$ is expected to
be the dominant decay mode with
a branching fraction somewhere between $40\sim 90$\%, 
and the $h_c$ total width is expected to be less than 1~MeV~\cite{kuang}.

Three detection \& analysis methods have been used to study $h_{\, c}$ production and decay,
all of them use the processes indicated by arrows in 
Fig.~\ref{fig:charmonium}:
\begin{description}

\item[inclusive] In the ``inclusive'' mode, only the $\piz$ is detected
and the $h_{\, c}$ shows up as a peak in the mass recoiling against the
detected $\piz$, which is inferred from conservation of energy and momentum.
The inclusive mode signal yield is proportional the 
$Bf(\psip\rt\piz h_c)$.  This mode has the highest background.

\item[E1-tagged] In the ``E1-tagged'' mode the $\piz$ and the
E1 transition $\gamma$ from the $h_c\rt \gamma \etac$,
with energy in the range $465-535$~MeV, are detected.
The E1-tagged signal yield is proportional to the
branching fraction product 
$Bf(\psip\rt\gamma h_c)\times Bf(h_c\rt\gamma\etac)$.  
The background for this mode is relatively smaller than
that for the inclusive mode.

\item[exclusive] In the ``exclusive'' mode, the $\piz$, E1-$\gamma$
and all of the decay products of the $\etac$ are detected.  Here
all final-state particles are detected and energy-momentum
conserving kinematic fits can be used to improve the resolution.
The backgrounds are small and the yield is proportional
to a triple product of branching fractions, including that
for the $\eta_{\; c}$ decay channel that is detected.

\end{description} 
\noindent
The CLEO observation used both the E1-tagged and exclusive modes.
BESII has reported results from the inclusive and E1-tagged modes;
an exhaustive study of exclusive channels is in progress.

\begin{figure}
\begin{center}
\includegraphics[height=0.7\textwidth,width=0.7\textwidth]{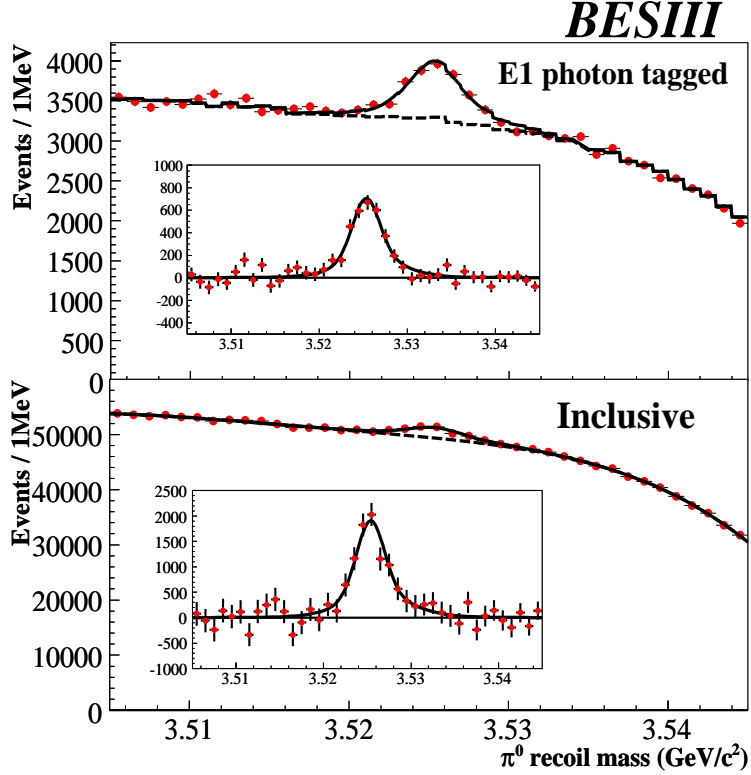}
\caption{The $\piz$ recoil mass for E1-tagged (top) 
and inclusive (bottom)$\psip\rt\piz X$
decays.  The insets show the signal yields with the fitted backgrounds
subtracted. 
} 
\label{fig:hc}
\end{center}
\end{figure}

The BESIII $\piz$ recoil mass distributions for the E1-tagged and
inclusive modes are shown in Fig.~\ref{fig:hc}.  The E1-tagged 
sample (top) has the most distinct signal and this
is used to determine the mass and width of the $h_c$.
The solid curve in the figure is the result of a fit using a BW
function convolved with a MC-determined resolution function to
represent the signal, and a background shape that is determined
from events with no photon in the E1 signal region, but with a
photon in the E1-tag sidebands.  From the fit, the mass and
width are determined to be     
\begin{eqnarray}
m_{h_c} &=& 3525.40 \pm 0.22~{\rm MeV}\\
\Gamma_{h_c} &=& 0.73 \pm 0.53~{\rm MeV};
\end{eqnarray}
the 90\% CL upper limit on the width is $\Gamma_{h_c}<1.44$~MeV.
With this mass value, the $P$-wave hyperfine splitting is 
$<m_{\chi_{cJ}}> - m_{h_c}=-0.10\pm 0.22$~MeV, consistent with zero.
From the signal yield, the product branching fraction
$Bf(\psip\rt\piz h_c)\times Bf(h_c\rt\gamma\etac) = (4.48\pm 0.64)\times 10^{-4}$
is determined.

The inclusive $\piz$ recoil mass distribution is shown in the
lower part of Fig.~\ref{fig:hc}. Here the solid curve is the result
of a fit where the mass and width of the signal function are
fixed at the E1-tagged results and the background is parameterized
by a fourth-order Chebyshev polynomial with all parameters allowed 
to float.  The signal yield and the product branching fraction results
from the E1-tagged mode
are used to make the first determination of the individual branching fractions:
\begin{eqnarray}
Bf(\psip\rt\piz h_c)   &=& (8.4 \pm 1.6)\times 10^{-4}\\
Bf(h_c\rt\gamma\etac ) &=& (54.3 \pm 8.5)\% ,
\end{eqnarray}
which are within the range of theoretical expectations.

\section{Other BESIII activities}

The results described in this report correspond to only a small fraction
of the total BESIII physics activity.  A more complete overview of
the entire physics program planned for BESIII is provided in an 800 page
tome published in 2009~\cite{BESIII-book}.

\subsection{Charmed meson physics}

\begin{figure}
\begin{center}

\mbox{
  \includegraphics[height=0.4\textwidth,width=0.3\textwidth,angle=270]{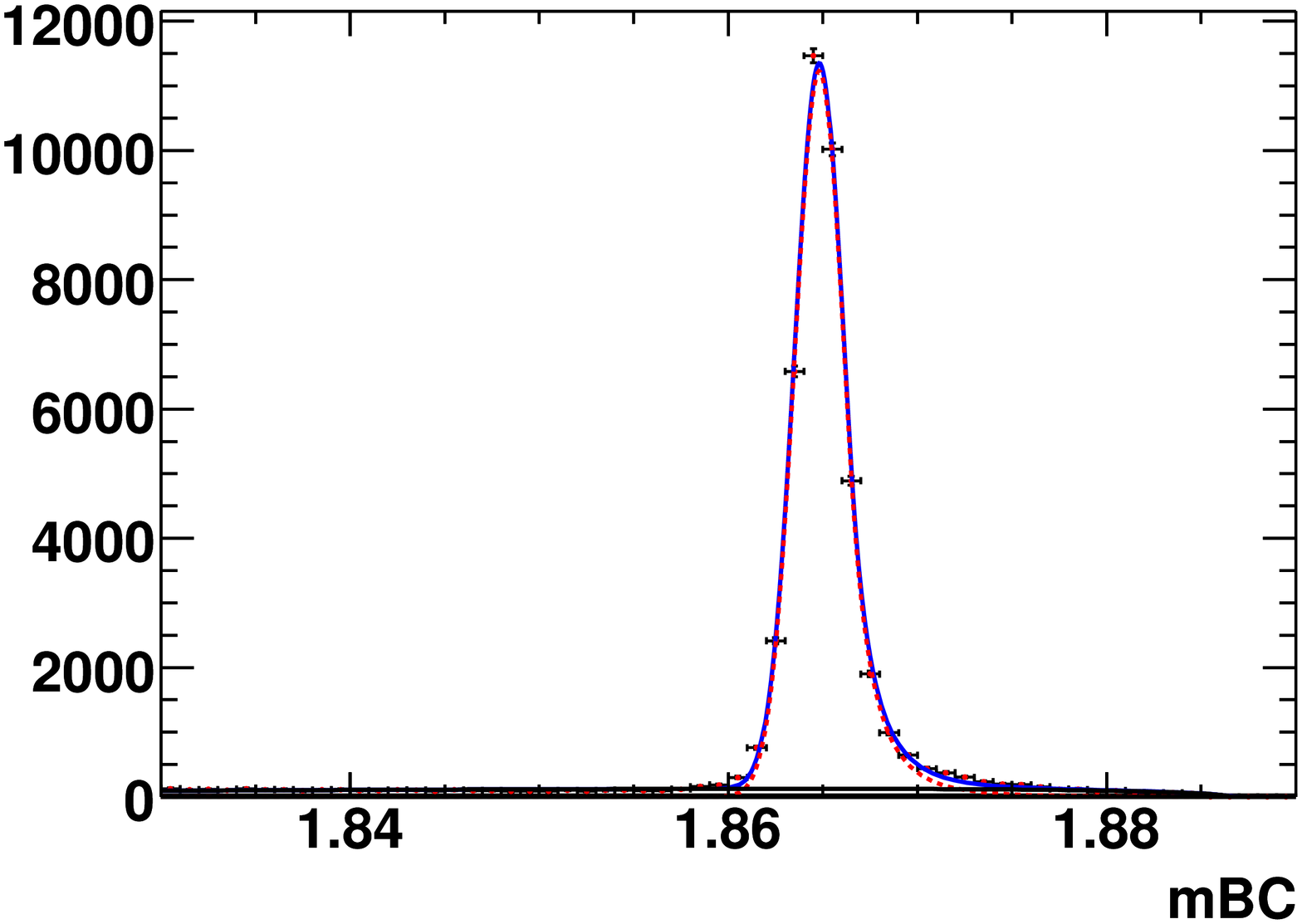}
}
\mbox{
  \includegraphics[height=0.4\textwidth,width=0.3\textwidth,angle=270]{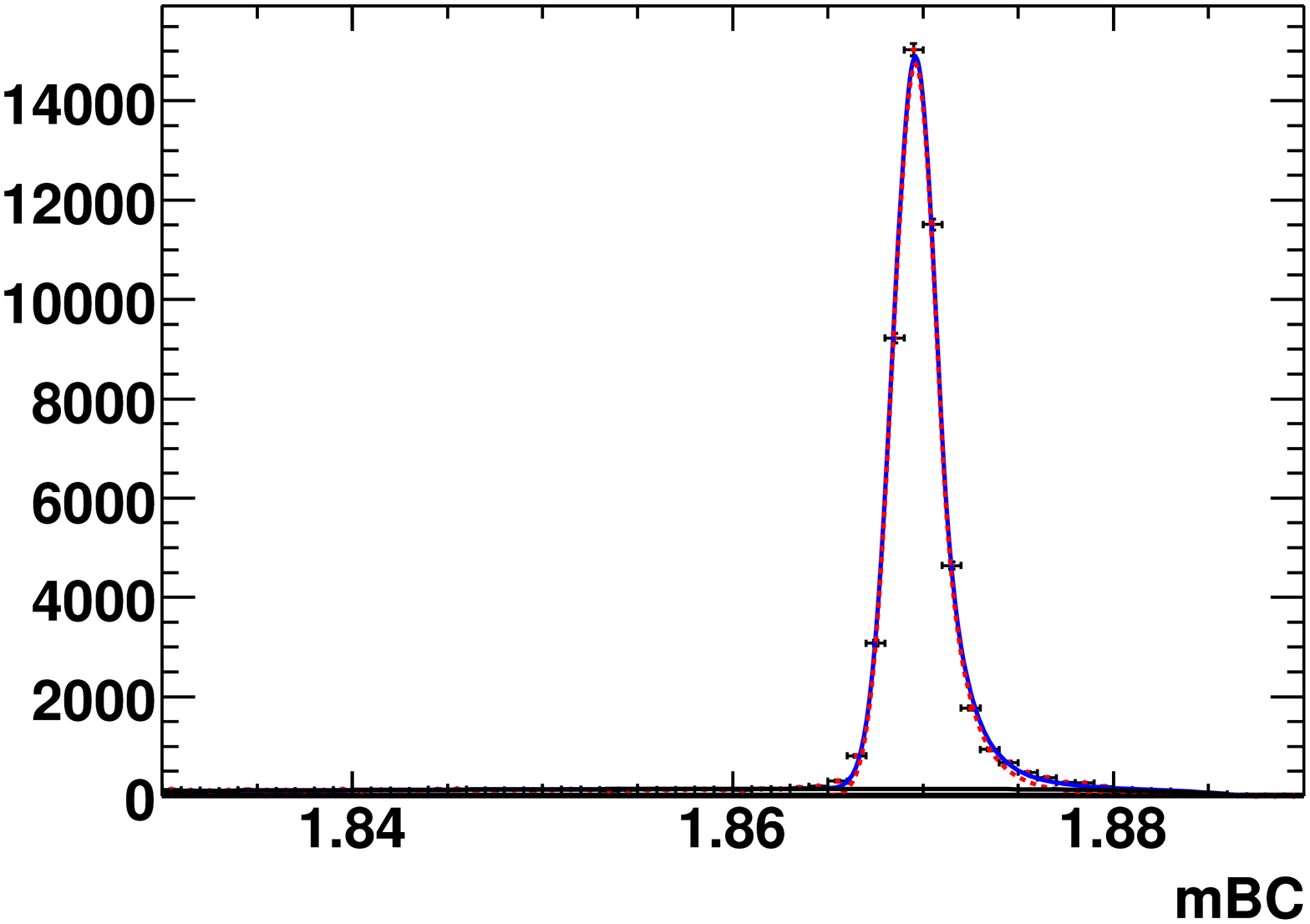}
}
\mbox{
  \includegraphics[height=0.4\textwidth,width=0.3\textwidth,angle=270]{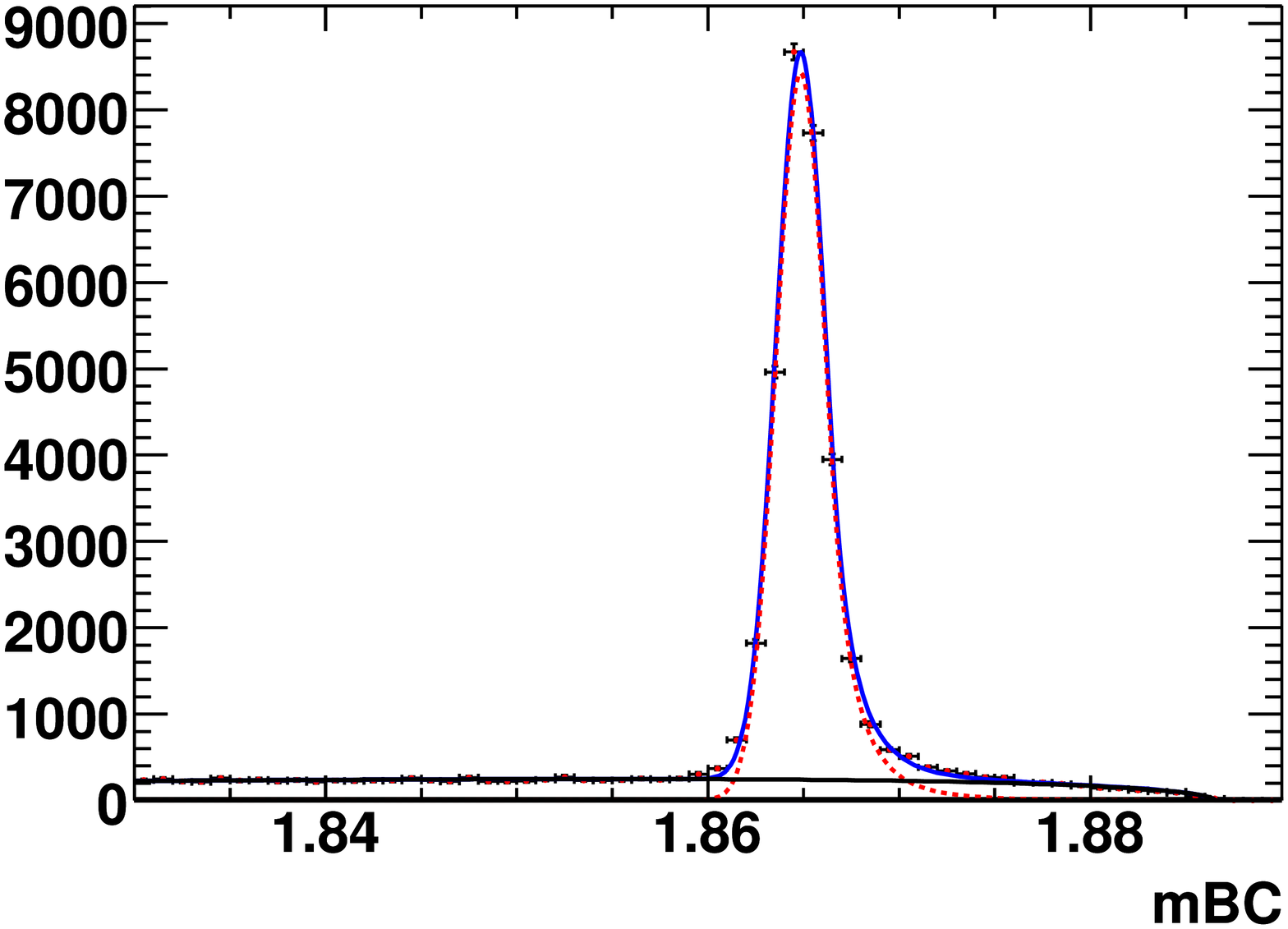}
}
\mbox{
  \includegraphics[height=0.4\textwidth,width=0.3\textwidth,angle=270]{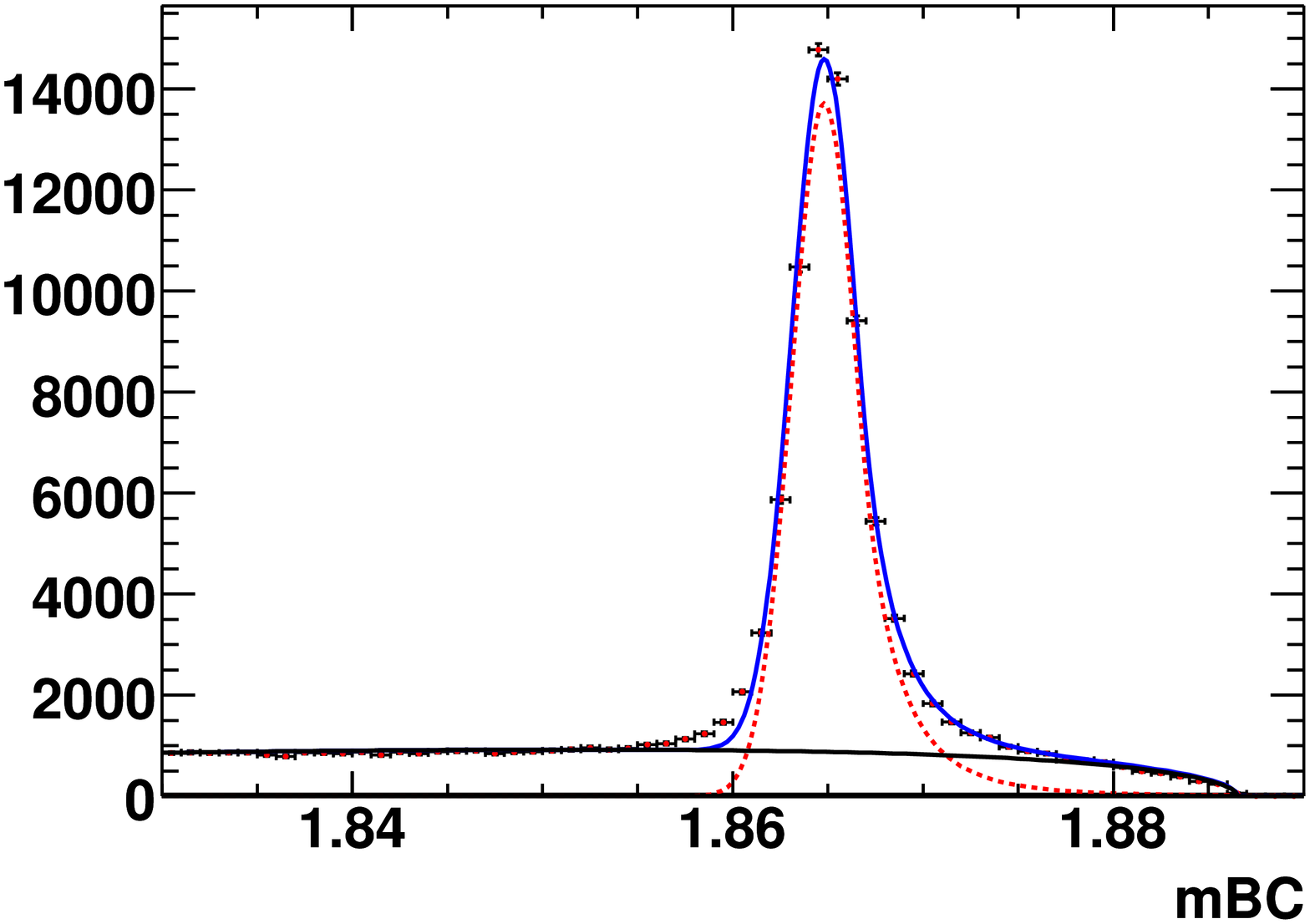}
}
\caption{Reconstructed $D$-meson mass distributions for {\bf top:}
$D^0\rt K^-\pi^+$ (left) \& $D^+\rt K^-\pi^+\pi^+$ (right), and {\bf bottom:}
$D^0\rt K^-\pi^+\pipi$ (left) \& $D^0\rt K^-\pi^+\piz$ (right).
The horizonal axis units are GeV.
} 
\label{fig:mbc} 
\end{center}
\end{figure}

The most notable subject that I have skipped in this report is the  BESIII
charmed physics program that is aimed primarily at precision studies of weak decay
processes of $D$ and $D_s$ mesons.  The initial phase of this program was
a long data-taking run that accumulated 2.9~fb$^{-1}$ 
at the peak of the $\psi(3770)$ charmonium meson.
This is a resonance in the $\ee\rt \DDbar$ channel with a peak cross section
of about 6~nb at a c.m. energy that is about 40~MeV
above the $E_{c.m.}= 2m_{\, D}$ open-charm mass threshold.  (The $\psi(3770)$
is included in the sketch of the charmonium spectrum shown in Fig.~\ref{fig:charmonium}.)
At least 90\%, maybe more, of $\psi(3770)$ decays are to $\DDbar$ meson pairs and nothing else;
there is not enough enough c.m. energy to produce any other accompanying hadrons.  
As a result, The energy of each $D$ meson is half of the total c.m. energy, which
is precisely known.  Thus, when a $D$ meson is reconstructed in an event,
the recoil system is ``tagged'' as a~$\bar{D}$, and the
constraint on the energy results in reconstructed $D$-meson mass signals
that have excellent resolution ($\sigma= 1.3$~MeV for all charged modes
and $\sigma=1.9$~MeV for modes with one $\piz$) and signal to noise.  $D$-meson
signals for four commonly used tag decay modes are shown in Fig.~\ref{fig:mbc}. 
Moreover, the $\DDbar$ system is in a coherent, $P$-wave
quantum state with $J^{PC}=1^{--}$.  This coherence is unique to $D$ mesons
originating from $\psi(3770)\rt \DDbar$ decays and
permits a number of interesting measurements~\cite{asner}.  For example,
if one $D$ meson is tagged in a pure $CP$ decay mode (such as $K^+K^-$, $\pipi$
or $K_S\piz\piz$ for $CP=+1$, and  $K_S\piz$, $K_S\eta$ or $K_S\omega$ for
$CP=-1$), the decay of the accompanying $D$ meson to a $CP$ eigenstate with the same
$CP$ eigenvalue would be an unambiguous signal for $CP$ violation.

The 2.9~fb$^{-1}$ data sample that has already been collected
contains almost 20M $\DDbar$ meson pairs and
is about three times the world's previous previously largest $\psi(3770)$
event sample --collected by CLEOc-- and is currently being used for numerous
analyses aimed at searches for rare decays \& new physics, and improving on the precision
of previous measurements.  In many of the latter cases, the measurements are of
form-factors that are accessible in lattice QCD
calculations.  As the precision of lattice QCD improves, BESIII will provide
more precise measurements that continue to challenge the theory.  

Ultimately, over the next seven years or so, BESIII intends to collect a total
of $\sim 10$~fb$^{-1}$ at the $\psi(3770)$ for $D$ meson measurements
and a comparable sample at higher energy for $D_s$ meson studies.

\subsection{Additional $\jpsi$ and $\psip$ data samples.}

The results reported above are based on 106M event $\psip$ and 225M event
$\jpsi$ data samples.  The ultimate goal of the BESIII program is to collect
a total of $\sim 10^{\, 9}$ $\psip$ events and a multiple of $10^{\, 9}$ $\jpsi$
events.  These samples will be used, among other things, for detailed PWA
of the many  unassigned resonance peaks that have been seen, studies of baryon
spectroscopy, and high-statistics measurements of isospin-violating processes
that are proving to be valuable probes of the structure of near-threshold
resonances.  In addition, with the huge $\jpsi$ data sample, the
expected SM level for weak decays of the $\jpsi$ to final states containing
a single $D$ or $D_s$ meson can be accessed and searches for
non-SM weak decays and lepton-flavor-violating decays, such as $\jpsi\rt e^+\mu^-$,
will have interesting sensitivity.

\subsection{QCD, two-photon, $\tau$ and $XYZ$-meson physics}

BESIII also plans to redo the total cross section measurements for $\ee\rt hadrons$
with higher precision over the entire accessible c.m. energy range,
measure $\piz$ and $\eta$ formfactors in two-photon collisions, remeasure
the $\tau$ mass with much improved accuracy, and do studies of the recently
discovered $XYZ$ mesons.
   
Cross section measurement scans will cover c.m. energies from near the
nucleon-antinucleon threshold up to the $\Lambda^+_c \Lambda^-_c$
threshold.  The data near the nucleon-antinucleon threshold will be used 
to measure neutron form factors~\cite{baldini}.  

The $\tau$ mass measurement will benefit from a high-precision
beam-energy monitor based on the detection of Compton scattering
of back-scattered photons from a high powered single-mode
infrared laser.  This system has been commissioned and
routinely measures the beam energy with a precision
of $\delta_{E_{\rm beam}}/E_{\rm beam} \simeq 10^{-5}$~\cite{bes3_bem}.

Data taken in a dedicated run at $E_{c.m.}\simeq 4260$~MeV will
be used to study $Y(4260)$ decays.  Sensitive
searches for possible new, exotic mesons that decay to $\pi^+\jpsi$
and $\pi^+ h_c$, analogous to the $Z_1 (10610)^+$
and $Z_2(10650)^+$ mesons seen by Belle in the $\bbbar$
bottomonium meson system~\cite{belle_Z_b}, will be performed for
$\pipi\jpsi$ and $\pipi h_c$ final states. 

\section{Concluding remarks}

The BESIII experiment at the Institute of High Energy Physics
in Beijing, China is up and running and producing interesting
results on a variety fo topics.  The BEPCII collider is performing
near design levels and the BESIII detector performance is excellent.
We expect to produce many interesting new results in the coming decade.  

\section{Aknowledgements}
I thank the meeting organizers for inviting me to report on BESIII results 
at this interesting and well organized meeting.  I also thank
my BESIII colleagues for allowing me to represent them and
for generating the results discussed herein.   I benefited from
an informative conversation with Qiang Zhao;
Nik Berger, Liaoyuan Dong \& Yadi Wang provided me with some of the figures
used in this report; and Haibo Li provided a number of suggestions and corrections. 
This work was supported by
the National Research Foundation of Korea under 
WCU-program Contract No. R32-2008-000-10155-0 and
Grant No. 2011-0029457.

\end{document}